\documentclass[prl,aps,twocolumn,footinbib,superscriptaddress,notitlepage]{revtex4-1}
\usepackage{amsmath}
\usepackage{amssymb}
\usepackage{graphicx}
\usepackage{dcolumn}
\usepackage{enumerate}
\usepackage{bm}
\usepackage{xcolor}
\usepackage{comment}
\usepackage{subfigure}
\usepackage{breqn}
\usepackage{hyperref}
\hypersetup{colorlinks=true, citecolor=blue, urlcolor=blue, linkcolor=blue}

\makeatletter
\let\cat@comma@active\@empty
\makeatother

\begin{document}
\title{Quantum Enhanced Cavity QED Interferometer with Partially Delocalized Atoms in Lattices}
\author{Anjun Chu}
\email{anjun.chu@colorado.edu}
\affiliation{JILA, NIST and Department of Physics, University of Colorado, Boulder, Colorado 80309, USA}
\affiliation{Center for Theory of Quantum Matter, University of Colorado, Boulder, Colorado 80309, USA}
\author{Peiru He}
\affiliation{JILA, NIST and Department of Physics, University of Colorado, Boulder, Colorado 80309, USA}
\affiliation{Center for Theory of Quantum Matter, University of Colorado, Boulder, Colorado 80309, USA}
\author{James K. Thompson}
\affiliation{JILA, NIST and Department of Physics, University of Colorado, Boulder, Colorado 80309, USA}
\author{Ana Maria Rey}
\affiliation{JILA, NIST and Department of Physics, University of Colorado, Boulder, Colorado 80309, USA}
\affiliation{Center for Theory of Quantum Matter, University of Colorado, Boulder, Colorado 80309, USA}
\date{\today}

\begin{abstract}
We propose a quantum enhanced interferometric protocol for gravimetry and force sensing using cold atoms in an optical lattice supported by a standing-wave cavity.
By loading the atoms in partially delocalized Wannier-Stark states, it is possible to cancel the undesirable inhomogeneities arising from the mismatch between the lattice and cavity fields and to generate spin squeezed states via a uniform one-axis twisting model. 
The quantum enhanced sensitivity of the states is combined with the subsequent application of a compound pulse sequence that allows to separate atoms by several lattice sites. 
This, together with the capability to load small atomic clouds in the lattice at micrometric distances from a surface, make our setup ideal for sensing short-range forces.
We show that for arrays of $10^4$ atoms, our protocol can reduce the required averaging time by a factor of $10$ compared to unentangled lattice-based interferometers after accounting for primary sources of decoherence.
\end{abstract}

\maketitle

{\it Introduction.}---Ultracold atomic systems offer tremendous potential for quantum sensing applications including time keeping \cite{Ludlow2015} and gravimetry \cite{tino2020testing}, and thus provide  opportunities for searching or constraining new physics in outstandingly precise and compact experiments.
Despite the great advances in quantum sensing accomplished by cold atom experiments, one of the most important milestones that needs to be accomplished is to introduce quantum entanglement to enhance the  sensitivity of real-world sensors beyond the so-called standard quantum limit (SQL) attainable with uncorrelated particles \cite{tse2019quantum,acernese2019increasing,Malnou2019,backes2021quantum}.

Important steps towards this goal have been accomplished such as the generation of up to $19$~dB spin squeezing in cavities \cite{Hosten2016,Cox2016,Schleier-Smith2010,Vladan2020}.
Nevertheless, the use of entangled states in state-of-the-art inertial sensors has yet to be achieved. 
Limitations include the spatial mismatch between the lattice potential and the cavity mode which degrades the utility of spin squeezing after releasing the atomic cloud to free space \cite{Hu2015}. 
Conventional free-falling experiments also lack spatial resolution and suffer from limited interrogation time \cite{alauze2018trapped}.
Theoretical and experimental progresses to overcome these challenges have been reported in recent years although in different setups. 
For example, homogeneous atom-cavity couplings have been engineered by the use of commensurate lattices \cite{Hosten2016,Lee2014,Wu2019}, ring cavities \cite{bernon2011heterodyne,salvi2018squeezing,shankar2019squeezed,gietka2019supersolid}, or via time averaging as atoms free fall along the cavity axis \cite{cox2016spatially}.  
In parallel, lattice-based interferometers enjoying compact spatial volumes \cite{Tarallo2014,kessler2016situ,alauze2018trapped,Xu2019,morel2020determination} have reported capabilities to trap atoms near surfaces, and have achieved up to $20$~s holding time using uncorrelated atoms \cite{Xu2019}.

Here we propose a quantum enhanced lattice-based protocol that uses the motional eigenstates of the combined lattice plus gravity potential, the so-called Wannier-Stark (WS) states, to overcome relevant limitations faced by current atomic sensors. 
The key idea is the use of delocalized WS states over a few lattice sites, which enables averaging out the inhomegeneities of atom-cavity couplings at specific lattice depths.
This allows for the generation of uniform spin squeezed states via dynamical one-axis twisting (OAT) evolution \cite{kitawaga1993,wineland1992}, or via homogeneous quantum  nondemolition (QND) measurements \cite{Braginsky1996,Mandel1998}, even in incommensurate lattice-cavity geometries.
The uniformly generated spin squeezing are not only useful for quantum enhanced measurements of gravity \cite{Wu2019}, but also ideal for fundamental tests of short-ranged forces \cite{alauze2018trapped,Xu2019} which require loading small atomic clouds close to a surface or a source mass, such as dark energy \cite{Matt2017}, Casimir-Polder forces \cite{Harber2005}, and non-Newtonian corrections of gravity \cite{Kapner2007,Sushkov2011}. Furthermore, the ability to tune the inhomogeneities of couplings to a bosonic mode that mediates interactions or introduces disorder opens new possibilities in quantum many-body simulators \cite{Ritsch2013}. 

Our work focuses on the dynamical generation of spin squeezing and the interferometric sequence to transfer the atoms to WS orbitals separated by several lattice sites to improve phase accumulation. 
Moreover, the interferometric phase can be mapped into a magnified rotation of the atomic internal state by reversing the squeezing protocol, which can be measured without the need of below-SQL detection resolution \cite{Davis2016,hosten2016quantum,Schulte2020ramsey}.
After accounting for primary sources of decoherence, we show that applying our scheme to arrays of $10^4$ atoms, it should be possible to detect short-range forces acting at $\mu$m-scale distances, with an averaging time reduced by a factor of $10$ compared to unentangled lattice-based interferometers \cite{Tarallo2014,alauze2018trapped}.

\begin{figure}[t]
    \centering
    \includegraphics[width=6cm]{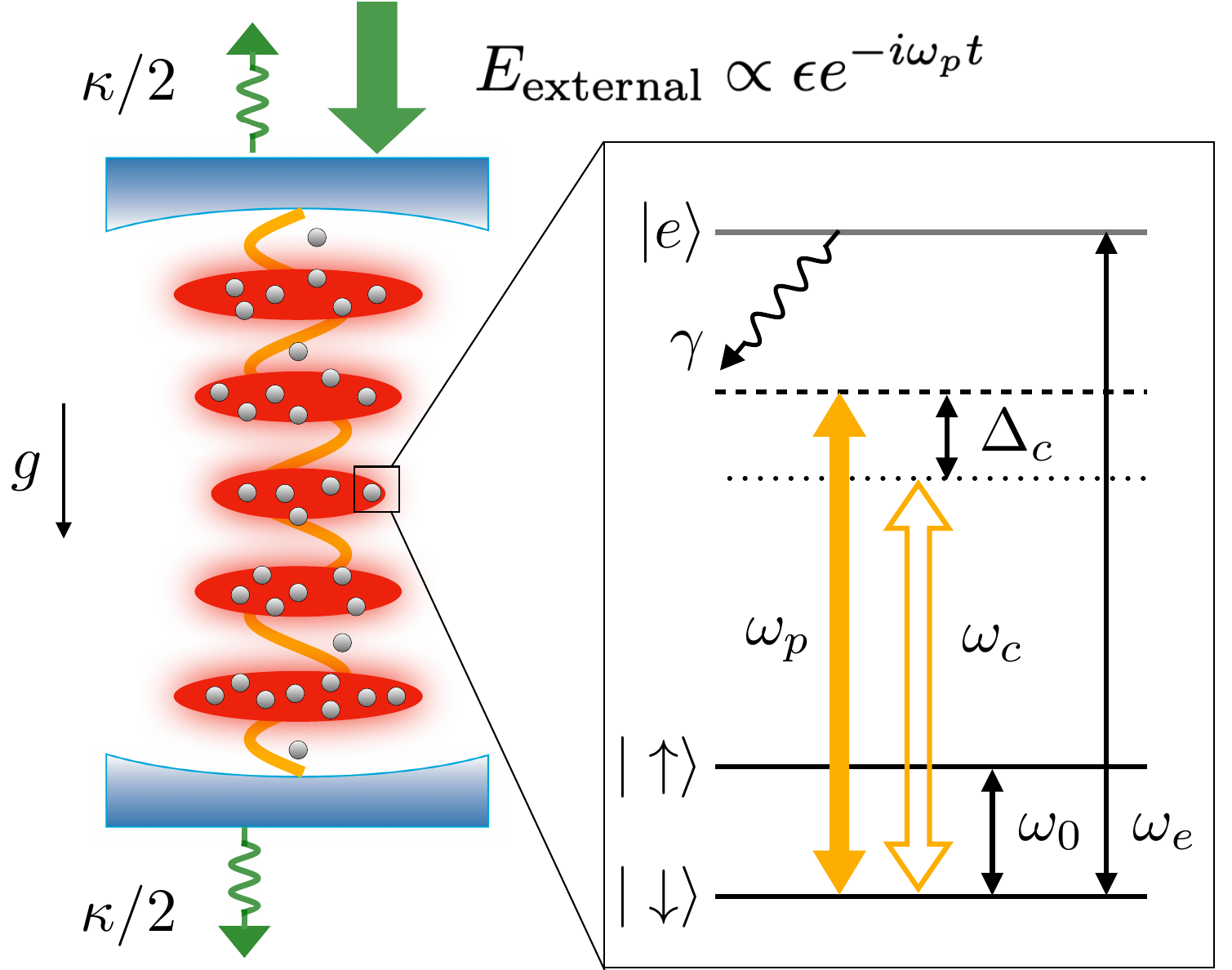}
    \caption{\label{fig1} Protocol schematics: An ensemble of ultracold atoms are trapped in the lowest band of a lattice supported by a standing-wave optical cavity oriented along the direction of the gravitational acceleration $g$. The cavity decay rate is $\kappa/2$ on each side. The two long-lived internal levels of an atom with energy splitting $\hbar\omega_0$ act as a spin-$1/2$ degree of freedom labeled as $|\uparrow\rangle$ and $|\downarrow\rangle$. These two states are coupled through a single cavity mode to the excited state $|e\rangle$, with energy $\hbar\omega_e$ and spontaneous emission rate $\gamma$. The cavity mode is coherently pumped by an external light field with detuning $\Delta_c=\omega_p-\omega_c$ from the cavity resonance which generates a net injected field in the cavity with amplitude $\epsilon$.}
\end{figure}

{\it System.}---We consider an ensemble of $N$ ultracold atoms, with mass $M$ trapped in a vertical standing-wave optical cavity, as depicted in Fig.~\ref{fig1}.
The atoms are confined in the lowest band of a one dimensional (1D) optical lattice oriented along the vertical direction $\hat{z}$. 
The gravitational potential with local acceleration $g$ generates a differential energy shift $Mgz$ between two atoms separated by a vertical distance $z$.  
Two long-lived internal levels of the atoms, with energy splitting $\hbar\omega_0$, are used to encode a spin-$1/2$ degree of freedom with states labeled as $|\uparrow\rangle$ and $|\downarrow\rangle$.  
A single cavity mode with frequency $\omega_c$ and wavelength $\lambda_c$ couples the $|\uparrow\rangle$ and $|\downarrow\rangle$ states to an optically excited state $|e\rangle$ of the atoms separated by a frequency $\omega_e$ from the $|\downarrow\rangle$ state. 
The atom-cavity coupling has a spatial profile $\mathcal{G}_{\uparrow,\downarrow}(z)=\mathcal{G}_{\uparrow,\downarrow}^{0}\cos(k_c z)$, where $k_c=2\pi/\lambda_c$. 
The cavity mode is coherently pumped by an external field detuned from the cavity resonance by $\Delta_c=\omega_p-\omega_c$.

We are focusing on the system operating in the dispersive regime of atom-light interaction, where both the pump and cavity mode are far-detuned from the atomic resonances, i.e. 
$\Delta_{\uparrow,\downarrow}\gg \mathcal{G}_{\uparrow,\downarrow}^{0}\sqrt{\langle\hat{a}^{\dag}\hat{a}\rangle}$, with $\Delta_{\uparrow}=\omega_p-\omega_{e}+\omega_0$ and $\Delta_{\downarrow}=\omega_p-\omega_{e}$.  
In this limit, the atomic excited state $|e\rangle$ can be adiabatically eliminated \footnote{See Supplemental Material at [URL will be inserted by publisher] for details of adiabatic elimination, model validity and experimental considerations, includes Ref.~\cite{james2007,Maschler2005,Prasanna2009,gluck2002,reiter2012,cox2016spatially,kolkowitz2017,Ji2013,FossFeig2013,Davis2016,Huelga1997}}, leading to the following Hamiltonian written in second quantized form in the rotating frame of the external optical pumping field, 
\begin{equation}
    \begin{aligned}
    \hat{H}&=\sum_{\beta=\uparrow,\downarrow}\int dz\hat{\psi}^{\dag}_{\beta}(z)\bigg[\frac{\hat{p}^2}{2M}+V_0\sin^2(k_lz)+Mgz\\
    &+\frac{\hbar|\mathcal{G}_{\beta}(z)|^2}{\Delta_{\beta}}\hat{a}^{\dag}\hat{a}\bigg]\hat{\psi}_{\beta}(z)+\hat{H}_{\mathrm{cav}}+\hat{H}_{\mathrm{drive}}.\\
    \end{aligned}
    \label{eq:hamil}
\end{equation}
Here, $V_0$ is the lattice depth, $k_l=2\pi/\lambda_l$ is the wavenumber of lattice beams that sets the atomic recoil energy $E_R=\hbar^2k_l^2/2M$ and the lattice spacing $a_l=\lambda_l/2$, where $\lambda_l$ is the wavelength of the lattice. 
The operator $\hat{a}$ is the annihilation field operator for cavity photons, and the operator $\hat{\psi}_{\beta}(z)$ annihilates an atom of spin $\beta$ at position $z$.
The cavity Hamiltonian is given by $\hat{H}_{\mathrm{cav}}/\hbar=-\Delta_c\hat{a}^{\dag}\hat{a}+\varepsilon \hat{a}^{\dag}+\varepsilon^{*}\hat{a}$, where $\varepsilon$ is the amplitude of the injected field. 
The drive Hamiltonian $\hat{H}_{\mathrm{drive}}/\hbar=\int dz[\Omega\hat{\psi}_{\uparrow}^{\dag}(z)\hat{\psi}_{\downarrow}(z)+\mathrm{h.c.}]-\delta\hat{\psi}_{\uparrow}^{\dag}(z)\hat{\psi}_{\uparrow}(z)$ describes a switchable external microwave drive, with Rabi frequency $\Omega$, drive detuning $\delta$ that uniformly couples the spin-$1/2$ degree of freedom when applied.
 
We expand the atom field operators $\hat{\psi}_{\beta}(z)$ in terms of the Wannier-Stark (WS) orbitals: $\hat{\psi}_{\beta}(z)=\sum_n \hat{c}_{n\beta}\phi_n(z)$, where $\hat{c}_{n\beta}$ annihilates an atom of spin $\beta$ in the WS state $|\phi_n\rangle$ centered at site $n$. 
In the tight-binding limit, the wave function of the WS state $|\phi_n\rangle$ takes the form $\phi_n(z)=\sum_m\mathcal{J}_{m-n}(2J_0/Mga_l)w(z-ma_l)$ \cite{gluck2002},
where $\mathcal{J}_n(x)$ is the Bessel function of the first kind, $J_0/\hbar$ is the nearest-neighbor tunneling rate, and $w(x)$ is the ground band Wannier function.
If we assume the cavity-induced AC Stark shifts $\hbar|\mathcal{G}^0_{\uparrow,\downarrow}|^2\langle\hat{a}^{\dag}\hat{a}\rangle/\Delta_{\uparrow,\downarrow}$ are smaller than $Mga_l$ \cite{Note1}, so transitions between WS orbitals are suppressed, the atom-cavity dynamics can be simplified into the following Hamiltonian,
\begin{equation}
   \hat{H}=\hbar\sum_n\Big(-\delta+\eta_n\hat{a}^{\dag}\hat{a}\Big)\hat{S}^z_n+H_{\mathrm{cav}}+\hbar\sum_n\Omega\hat{S}^x_n.
    \label{eq:qnd}
\end{equation}
Here, the spin operators are defined in terms of atomic creation and annihilation operators for $|\uparrow_n\rangle\equiv|\uparrow;\phi_n\rangle$ and $|\downarrow_n\rangle\equiv|\downarrow;\phi_n\rangle$ states, $\hat{S}^{x,y,z}_n=\sum_{\beta,\beta'}\hat{c}^{\dag}_{n\beta}\sigma^{x,y,z}_{\beta\beta'}\hat{c}_{n\beta'}$, where $\sigma^{x,y,z}_{\beta\beta'}$ are the matrix elements of the corresponding Pauli matrices and $\beta,\beta'\in \{\uparrow,\downarrow\}$. 
It is also convenient to define the collective spin operators $\hat{S}^{x,y,z}=\sum_{n}\hat{S}^{x,y,z}_n$ for later discussions.
The dispersive atom-light coupling $\eta_n=\eta_n^{\uparrow}-\eta_n^{\downarrow}$, with $\eta_n^{\uparrow,\downarrow}=\int dz|\mathcal{G}_{\uparrow,\downarrow}(z)\phi_n(z)|^2/\Delta_{\uparrow,\downarrow}$, can be evaluated analytically, 
\begin{equation}
    \eta_n=\eta\bigg[1+\mathcal{C}
    \mathcal{J}_{0}\bigg(\frac{4J_0}{Mga_l}\sin(\varphi/2)\bigg)\cos(n\varphi)\bigg],
\end{equation} 
where $\eta=\frac{1}{2}\big(|\mathcal{G}_{\uparrow}^0|^2/\Delta_{\uparrow}-|\mathcal{G}_{\downarrow}^0|^2/\Delta_{\downarrow}\big)$ is the mean value of $\eta_n$ over all possible $n$, $\varphi=2\pi \lambda_l/\lambda_c$, and $\mathcal{C}$ is a constant of order one \cite{Note1}.  
We also replace $\Delta_c$ by an effective cavity detuning $\tilde{\Delta}_c=\Delta_c-\sum_n N_n(\eta_n^{\uparrow}+\eta_n^{\downarrow})/2$ in $\hat{H}_{\mathrm{cav}}$, where $N_n$ is the total atom number in $|\uparrow_n\rangle$ and $|\downarrow_n\rangle$ states. 

The last step is to adiabatically eliminate the injected light field and intracavity fluctuations, possible in the limits $\tilde{\Delta}_c\gg \eta\alpha\sqrt{N},\kappa$ \cite{Note1}, where $\alpha=\varepsilon/(\tilde{\Delta}_c+i\kappa/2)$ is the steady-state value of the cavity field, and $\kappa$ is the cavity intensity decay rate. With these reasonable approximations, the system can well described by an effective Hamiltonian involving only the spins,
\begin{equation}
  \hat{H}_{\mathrm{eff}}/\hbar=-\sum_n(\delta-\eta_n|\alpha|^2)\hat{S}^z_n+\sum_{nm}\chi_{nm} \hat{S}^z_n\hat{S}^z_m+ \Omega\sum_n\hat{S}^x_n .
    \label{eq:oat}
\end{equation}
When the microwave drive is off, the Hamiltonian above is the so-called one-axis twisting (OAT) model, with $\chi_{nm}=\eta_n\eta_m|\alpha|^2\tilde{\Delta}_c/(\tilde{\Delta}_c^2+\kappa^2/4)$ the OAT interaction strength, which is an iconic model for the generation of spin squeezed states \cite{kitawaga1993,wineland1992}. 

\begin{figure}[t]
    \centering
    \includegraphics[width=7.5cm]{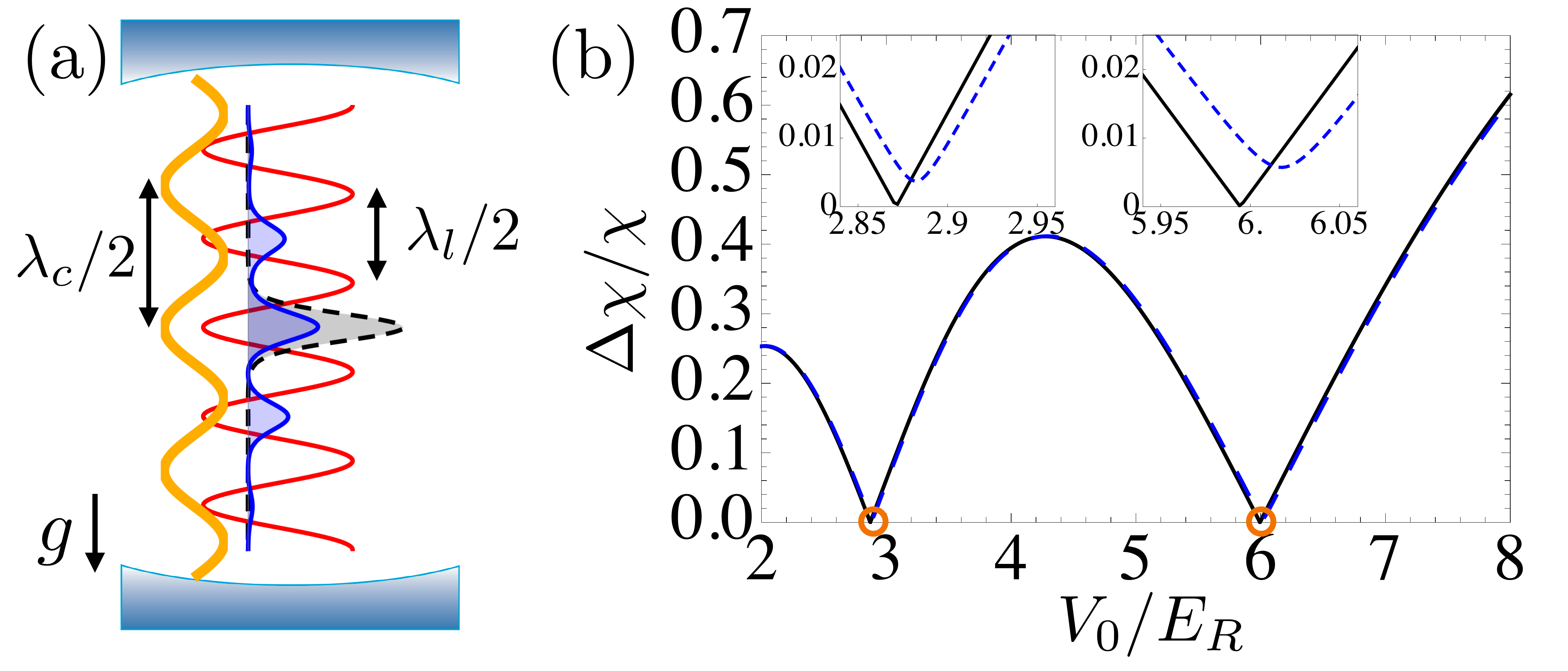}
    \caption{\label{fig2} (a) Inhomogeneous atom-light couplings arise due to the incommensurate wavelengths of the lattice beams (red curve) and the cavity mode (yellow curve), when atoms are frozen in Wannier states (black dashed curve) for deep lattice limit. The inhomogeneities can be cancelled out in a relatively shallow lattice since Wannier-Stark states (blue curve) can extend over a few lattice sites. (b) Standard deviation of the OAT coupling strengths $\Delta\chi=(\sum_{nm}(\chi_{nm}-\chi)^2/N)^{1/2}$ as a function of lattice depth $V_0/E_{R}$ assuming ${}^{87}$Rb atoms trapped in a $\lambda_l=532$~nm lattice. The black curve shows the magic lattice depths ($\Delta\chi=0$) can be achieved around $6.0E_R$ or $2.9E_R$ under ideal conditions, indicated by the orange circles. The blue dashed curve shows the imperfect cancellation of inhomogeneities in $\chi_{nm}$ with radial temperature $T=1\mu$K and radial trapping frequency $\omega_r/2\pi=1$kHz. The two insets show the zoomed $\Delta\chi$ near the magic lattice depths.}
\end{figure}

{\it Engineering homogeneous couplings.}---One limitation of spin squeezing generation protocols with frozen atoms in deep lattices ($J_0\approx 0$) is the inhomogeneous couplings arising from incommensurate lattice and cavity mode wavelengths ($\varphi\neq \pi j$ with $j$ an integer). 
However, in a relatively shallow lattice ($V_0<10E_R$) where $J_0\sim Mga_l$, the wave function of WS states can extend over a few adjacent lattice sites [see Fig.~\ref{fig2}(a)] due to non-negligible nearest-neighbor tunnel couplings, instead of being localized in a single site. The lattice depth can thus be used as a control  knob to vary the extension of the WS and for tuning  the inhomogeneitiy of the spin coupling parameters [see Fig.~\ref{fig2}(b)]. In particular, at the magic lattice condition, 
\begin{equation}
    \mathcal{J}_0\bigg(\frac{4J_0}{Mga_l}\sin(\varphi/2)\bigg)=0,
    \label{eq:magic}
\end{equation}
we can completely  average  out the inhomogeneities and  obtain uniform couplings in Eq.~(\ref{eq:oat}) with $\eta_n=\eta$ and
$\chi_{nm}=\chi=\eta^2|\alpha|^2\tilde{\Delta}_c/(\tilde{\Delta}_c^2+\kappa^2/4)$.
This technique is relevant  not only for the generation of  homogeneous spin squeezing  but also for  quantum simulation of long-range spin models with tunable inhomgeneity \cite{Ritsch2013}.
In practice, the thermal distribution of atoms in the radial direction and the undesirable couplings between axial and radial confinement of the Gaussian beam profile can lead to an imperfect cancellation [see the insets in Fig.~\ref{fig2}(b)], which can be highly suppressed by operating at low radial temperature or large radial confinement \cite{Note1}.
For $^{87}$Rb atoms with $\lambda_c=780$~nm ($\mathrm{D}_2$ transition) and $\lambda_l=532$~nm, the magic lattice depths are around $6.0E_R$ and $2.9E_R$ [see Fig.~\ref{fig2}(b)].
For $^{171}$Yb atoms with $\lambda_c=556$~nm ($^1\mathrm{S}_0\rightarrow {}^3\mathrm{P}_1$ transition) and $\lambda_l=413$~nm \cite{dzuba2010dynamic}, the magic lattice depth is around $3.2E_R$. The negligible scattering length of $^{171}$Yb atoms \cite{cazalilla2014ultracold} and their insensitivity to magnetic and electric fields make them ideal for inertial sensing. 
For the cases above, at the smallest magic lattice depth the WS state spreads within three lattice sites. 

\begin{figure}[t]
    \centering
    \includegraphics[width=7.5cm]{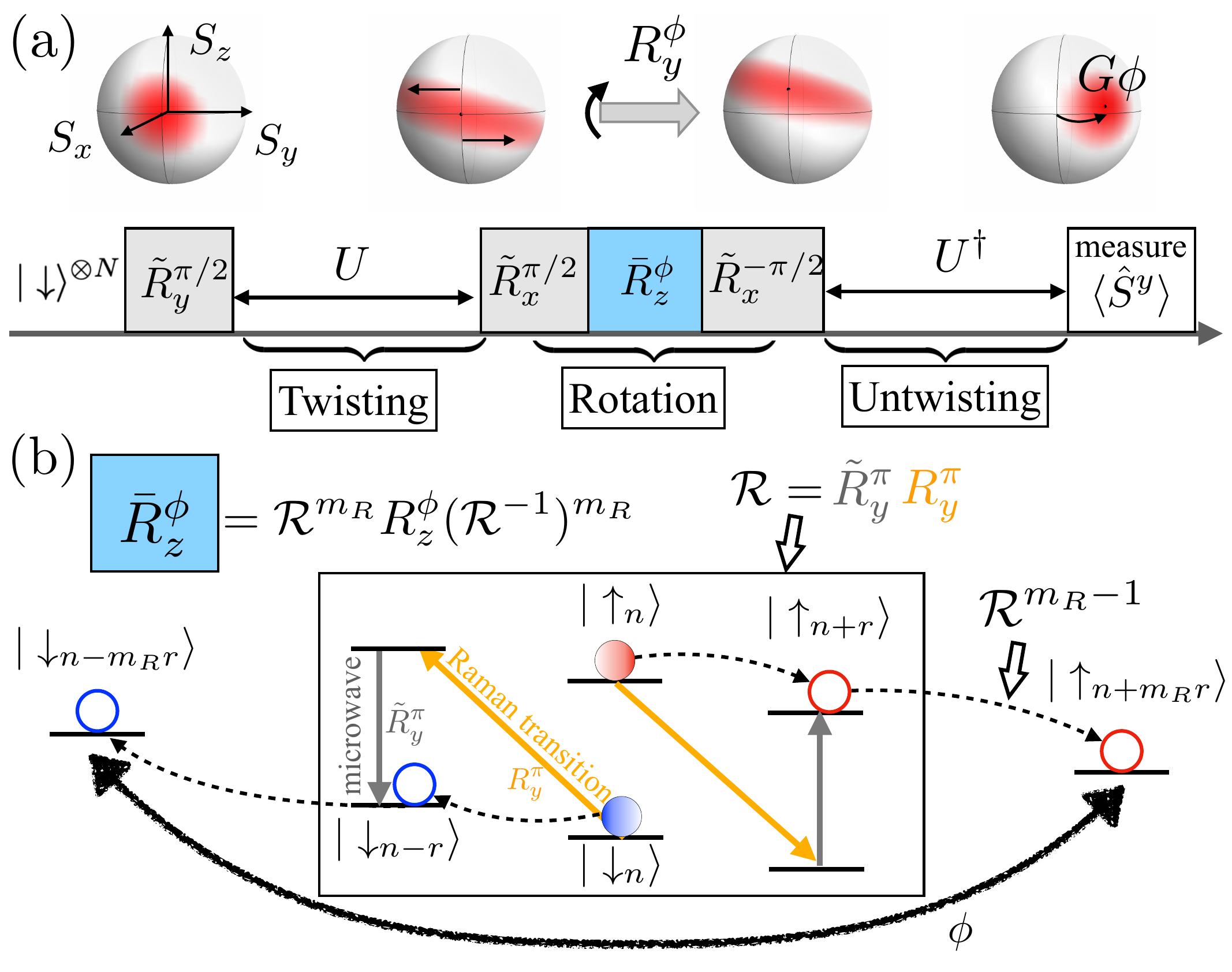}
    \caption{\label{fig3} Schematic of the quantum-enhanced gravimetry using Wannier-Stark (WS) states. (a) After the preparation of a coherent spin state along $\hat{x}$ direction in the carrier transition, we apply the twisting Hamiltonian for a time $t_0$ as $U=\exp(-i\chi\hat{S}^z\hat{S}^zt_0)$, and the system becomes a squeezed state sensitive to small rotations about the $\hat{y}$ axis ($\tilde{R}_y^{\phi}$). By applying the untwisting sequence $U^{\dag}$ for the same amount of time $t_0$, the quantum noise returns to the SQL level and the small rotation angle $\phi$ is amplified into a larger angle $G\phi$ around $\hat{z}$ axis, which can be detected by measuring $\langle \hat{S}^y\rangle$. (b) The phase accumulation due to gravitational energy difference is achieved through a compound pulse sequence that separates the atoms in the corresponding WS states by $2m_Rr$ lattice sites. A single compound pulse $\mathcal{R}$, as shown in the box, is a combination of a microwave pulse in the carrier transition and a Raman pulse for the $r$-th WS sidebands, which generates spin-dependent spatial transfer of the atoms (indicated by red/blue circles) from $|\uparrow_n\rangle/|\downarrow_n\rangle$ to $|\uparrow_{n+r}\rangle/|\downarrow_{n-r}\rangle$.}
\end{figure}

{\it Quantum enhanced interferometric protocol.}---Since the energy splitting of WS states is proportional to the gravitational acceleration $g$, our system can be directly used for quantum enhanced gravimetry. 
The protocol consists of the following steps, as illustrated in Fig.~\ref{fig3}. After the application of a short $\pi/2$ pulse with the microwave drive in an empty cavity [see Eq.~(\ref{eq:oat}) with $\alpha=0$] that prepares a spin coherent state along $\hat{x}$ direction, the system is let to evolve for a time $t_0$ under the OAT interaction mediated by the optical cavity [see Eq.~(\ref{eq:oat}) with $\Omega=0$ using an additional spin echo $\pi$ pulse at $t_0/2$ to cancel additional $\hat{S}^z$ rotations], which results in the generation of a uniform spin squeezed state [see Fig.~\ref{fig3}(a)]. 
The reduced noise quadrature of the state makes it highly sensitive to small rotations about the $\hat{y}$ axis, $\tilde{R}_y^{\phi}=\mathrm{e}^{-i\phi \hat{S}^y}$.

To perform precise measurement of a phase $\phi$ arising from gravitational energy shifts, Raman sideband transitions to WS states separated by a few lattice sites are used. Explicitly, the rotation about the $\hat{y}$ axis is implemented as $\tilde{R}_y^{\phi}=\tilde{R}_x^{-\pi/2}\bar{R}_z^{\phi}\tilde{R}_x^{\pi/2}$, with  $\bar{R}_z^{\phi}=(\mathcal{R}^{\dag})^{m_R}R_z^{\phi}\mathcal{R}^{m_R}$, where $m_R$ is the number of imposed compound pulses, and $\mathcal{R}=\tilde{R}^{\pi}_yR^{\pi}_y$ is a compound pulse to separate the atoms in $|\uparrow\rangle$ and $|\downarrow\rangle$ states by $2r$ lattice sites: $|\uparrow_n\rangle\rightarrow|\uparrow_{n+r}\rangle$ and $|\downarrow_n\rangle\rightarrow|\downarrow_{n-r}\rangle$. 
It consists of a $\pi$ Raman pulse $R^{\pi}_y$ with appropriate momentum kick and frequency $\omega_{\mathrm{R}}$ to perform the desired $r$-site transfer in the lattice (apply from side to ensure homogeneity for all atoms), followed by a $\pi$ pulse on the carrier transition to flip back the spin using a microwave drive ($\tilde{R}^{\pi}_y$) with frequency $\omega_{\mathrm{MW}}$ [see Fig.~\ref{fig3}(b)]. 
Such type of compound pulse sequences have been already successfully demonstrated in $^{87}$Rb atoms \cite{pelle2013state}. 
The $R_z^{\phi}$ operator describes the free evolution for a time $\tau$ of the atoms separated by $2m_R r$ lattice sites when they accumulate a phase  $\phi=(\omega_{\mathrm{R}}-\omega_{\mathrm{MW}}-Mga_lr/\hbar)\times 2m_R\tau$.
Note that one can apply an additional microwave pulse $\tilde{R}^{\pi}_y$ at $\tau/2$ to remove the undesirable hyperfine energy splitting $\hbar\omega_0$.

\begin{figure}[t]
    \centering
    \includegraphics[width=7.5cm]{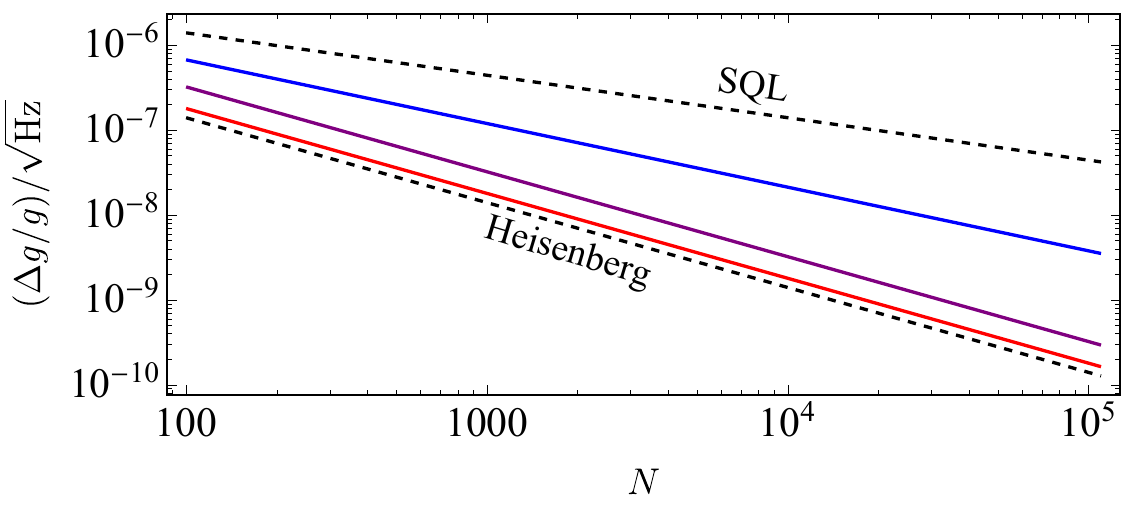}
    \caption{\label{fig4} Interferometer sensitivity $\Delta g/g$ as a function of atom number $N$, assuming $5.32~\mu$m separation for $^{87}$Rb atoms via compound pulse sequence, $1$~s phase accumulation time, and $C'=2$. The red curve indicates the ideal implementation without decoherence, while the blue curve and purple curve take account the effect of cavity loss and spontaneous emission with spin flip probability $P_f=1/2$ and $P_f=0$ respectively.}
\end{figure}

Finally, one can perform a time reversal of OAT dynamics by changing the frequency of pump laser such that $\tilde{\Delta}_c\rightarrow -\tilde{\Delta}_c$, followed by a measurement of $\langle \hat{S}^y\rangle$ \cite{Davis2016}.
Under this untwisting sequence, the accumulated phase $\phi$ is amplified by a factor of $G=(\partial_{\phi}\langle \hat{S}^y\rangle/S)_{\phi\rightarrow 0}$, and the quantum noise for phase measurement $\sigma_{p}=(\Delta S^y/S)_{\phi\rightarrow 0}$ returns to the SQL level, $(\sigma_p)_{\mathrm{SQL}}=1/\sqrt{N}$, which leads to a phase sensitivity $\Delta\phi=\sigma_p/G$ achievable with detection resolution at the atom shot noise level.
So we estimate a sensitivity of gravimetry by $\Delta g/g=\xi/(\phi_g\sqrt{N})\times \sqrt{\tau/T}$, where $\phi_g=2Mga_lrm_R\tau/\hbar$, $\xi^{-2}=1/[N(\Delta\phi)^2]$ is the metrological gain over the SQL, and $T$ is the averaging time.
The optimal sensitivity approaches the Heisenberg limit $\Delta g/g\propto 1/N$ under pure Hamiltonian dynamics [see the red curve in Fig.~\ref{fig4}]. 

Our protocol could be also ideal for sensing weak short-range forces generated by an object placed close to the atoms \cite{Matt2017,Harber2005,Kapner2007,Sushkov2011}, which introduces new possibilities in exploring new physics beyond the Standard Model. Such forces will generate an additional potential $\mathcal{U}(z)$ that will mainly modify the phase accumulated by an atom in the WS state centered at site $n$ to $\tilde{\phi}_n=\phi+(\mathcal{U}_{n+m_Rr}-\mathcal{U}_{n-m_Rr})\tau/\hbar$, where $\mathcal{U}_n=\int dz\,\mathcal{U}(z)|\phi_n(z)|^2$.
Given the dependence of the phase on initial WS states, which will dephase the atomic sample if spreading over multiple WS states, the use of atomic clouds with small spatial extension to reduce the number of occupied WS states can be crucial. 
For these situations, since the inhomogeneities in atom-light couplings do not average out in a single realization, one needs to account for important systematic errors in the amplification factor $G$, in contrast to the subdominant suppression of $G$ when the atomic array is fully spread across the lattice \cite{Note1}. 
Therefore, the magic lattice condition can lead to significant improvements if the atoms are restricted to local regions of the lattice.

{\it Experimental considerations.}---Experimental imperfections such as cavity loss and spontaneous emission of the excited state during the spin squeezing generation and other dephasing mechanisms during the interrogation will degrade the ideal sensitivity in practical implementations as we now discuss.
Cavity loss induces phase fluctuations of the collective spin with collective dephasing rate $\Gamma_z=\chi\kappa/\tilde{\Delta}_c$, which lead an increase in the variance of $\hat{S}^y$.
Spontaneous emission from the excited state $|e\rangle$, at a rate $\gamma$, generates off-resonant photon scattering processes with a total rate $\Gamma\propto \gamma|\mathcal{G}^0_{\uparrow,\downarrow}|^2|\alpha|^2/\Delta_{\uparrow,\downarrow}^2$, including single-particle spin flips and dephasing. 
Here we focus on the case with balanced spin flip rates, $\gamma_r=P_f\Gamma$ with $P_f$ the spin flip probability, which can be achieved by choosing appropriate energy levels and detunings, so the spontaneous emission generates no biases on the accumulated phase $\phi$.
The noise induced by the $\gamma_r$ terms is amplified during the untwisting protocol, making them the dominant single-particle noise source for measuring $\langle \hat{S}^y\rangle$.
The combination of cavity loss and spontaneous emission limits the metrological gain $\xi^{-2}$ to \cite{Note1},
\begin{equation}
    \xi^2\approx \frac{1+2N\Gamma_zt_0}{(N\chi t_0)^2}+\frac{8}{3}\gamma_rt_0,
\end{equation}
leading to an optimal value  $\xi^{-2}_{\mathrm{opt}}\propto\sqrt{NC'}$, where $C'=\chi^2/\Gamma_z\Gamma$ is related to the single-atom cooperativity \cite{Note1}.
This result translates into the sensitivity for gravimetry as $\Delta g/g\propto N^{-3/4}$ [see the blue curve in Fig.~\ref{fig4}]. Higher sensitivity can be reached by choosing specific schemes (e.g. cycling transitions) to suppress spin flip processes [see the purple curve in Fig.~\ref{fig4}].

Technical noise in experiment such as mechanical vibrations and local oscillator dephasing, as well as single particle decoherence due to interatomic interactions also impose a constraint on the interrogation time. 
In particular, single-particle decoherence imposes even more severe restrictions when operating with entangled states given their fragility to it \cite{Note1}.
For $^{87}$Rb assuming a $5.32~\mu$m atom separation achieved by $r=5$, $m_R=2$ in a $\lambda_l=532$~nm lattice, phase accumulation time $\tau=1$~s, $C'=2$, and spin flip probability $P_f=1/2$, one can achieve $\Delta g/g\sim 6\times 10^{-9}/\sqrt{\mathrm{Hz}}$ with about $5\times 10^4$ atoms, which is $20$~dB enhancement beyond SQL. 
If we compare this sensitivity with SQL for $\tau=10$~s, still a $10$~dB enhancement is possible, meaning that even after accounting for the fragility of the spin squeezed states, our protocol can not only reduce the required averaging time by a factor of $10$, but also increase the measurement bandwidth of time-varying signal by a factor of $10$, compared to unentangled lattice-based interferometers \cite{Tarallo2014,alauze2018trapped}.

{\it Conclusion and outlook.}---We proposed a quantum enhanced interferometric protocol using Wannier-Stark states in standing-wave cavity QED system, which allows for homogeneous spin squeezing generation and micrometric spatial resolution for gravimetry and force sensing. 
The many-body entanglement in our scheme leads to an order of magnitude reduction of the required averaging time compared to unentangled lattice-based interferometers. Our work opens new possibilities for quantum enhanced interferometry in versatile compact atomic sensors, as well as novel Hamiltonian engineering in quantum many-body simulators.

\begin{acknowledgments}
We thank John Robinson, Colin Kennedy, Matthew Affolter, Graham Greve, Chengyi Luo and Baochen Wu for useful discussions. 
This work is supported by the AFOSR Grant No. FA9550-18-1-0319, by the DARPA (funded via ARO) Grant No. W911NF-16-1-0576, the ARO single investigator Grant No. W911NF-19-1-0210, the NSF PHY1820885, NSF JILA-PFC PHY-1734006 and NSF QLCI-2016244 grants, by the DOE Quantum Systems Accelerator (QSA) grant and by NIST.
\end{acknowledgments}

\end{document}


\title{Quantum Enhanced Cavity QED Interferometer with Partially Delocalized Atoms in Lattices: Supplemental Materials}
\author{Anjun Chu}
\affiliation{JILA, NIST and Department of Physics, University of Colorado, Boulder, Colorado 80309, USA}
\affiliation{Center for Theory of Quantum Matter, University of Colorado, Boulder, Colorado 80309, USA}
\author{Peiru He}
\affiliation{JILA, NIST and Department of Physics, University of Colorado, Boulder, Colorado 80309, USA}
\affiliation{Center for Theory of Quantum Matter, University of Colorado, Boulder, Colorado 80309, USA}
\author{James K. Thompson}
\affiliation{JILA, NIST and Department of Physics, University of Colorado, Boulder, Colorado 80309, USA}
\author{Ana Maria Rey}
\affiliation{JILA, NIST and Department of Physics, University of Colorado, Boulder, Colorado 80309, USA}
\affiliation{Center for Theory of Quantum Matter, University of Colorado, Boulder, Colorado 80309, USA}
\date{\today}
\maketitle

\section{Cavity QED on Wannier-Stark states}
In the main text, we consider an ensemble of three-level atoms trapped in a standing-wave cavity along the vertical direction. 
A 1D optical lattice with lattice depth $V_0$ and wave vector $k_l=2\pi/\lambda_l$ along the cavity axis is used to confine the atoms.
Along this direction gravity imposes an additional linear potential of the form $Mgz$.
We assume that the spin-$1/2$ degree of freedom ($|\uparrow\rangle$, $|\downarrow\rangle$) is encoded in two long-lived internal states of the atoms with single particle energies $\{\hbar\omega_{\uparrow},\hbar\omega_{\downarrow}\}\equiv\{\hbar\omega_0,0\}$ respectively. 
They are coupled to an electronic excited state $|e\rangle$ (frequency $\omega_{e}$) by a single cavity mode with coupling strength $\mathcal{G}_{\uparrow,\downarrow}(z)=\mathcal{G}_{\uparrow,\downarrow}^{0}\cos(k_cz)$.
This cavity mode with resonant frequency $\omega_c$ is coherently pumped by an external field with intracavity intensity $\varepsilon$ and frequency $\omega_p$. 
We also include a homogeneous drive in the spin-$1/2$ degree of freedom with frequency $\omega_d$ and Rabi frequency $\Omega$. 
The Hamiltonian of the system can be written as the following second quantized form,
\begin{equation}
    \begin{aligned}
        H&=\sum_{\tau=\uparrow,\downarrow,e}\int\mathrm{d}z\psi^{\dag}_{\tau}(z)\bigg[\frac{p^2}{2M}+V_0\sin^2(k_lz)+Mgz+\hbar\omega_{\tau}\bigg]\psi_{\tau}(z)+\int dz\bigg[\hbar\mathcal{G}_{\uparrow}(z)a\psi^{\dag}_e(z)\psi_{\uparrow}(z)\\
        &+\hbar\mathcal{G}_{\downarrow}(z)a\psi^{\dag}_e(z)\psi_{\downarrow}(z)+\mathrm{h.c.}\bigg]+\int dz\bigg[\frac{\hbar\Omega}{2}\psi^{\dag}_{\uparrow}(z)\psi_{\downarrow}(z)\mathrm{e}^{-i\omega_d t}+\mathrm{h.c.}\bigg]+\hbar\omega_ca^{\dag}a+\hbar(\varepsilon\mathrm{e}^{-i\omega_pt} a^{\dag}+\varepsilon^{*}\mathrm{e}^{i\omega_pt}a).\\
    \end{aligned}
\end{equation}

It's convenient to transform the Hamiltonian into a frame rotating with the pump laser via the following unitary transformation,
\begin{equation}
    U=\exp\bigg\{-it\bigg[\int dz\bigg((\omega_p+\omega_{\downarrow})\psi^{\dag}_e(z)\psi_e(z)+(\omega_{\uparrow}+\delta)\psi^{\dag}_{\uparrow}(z)\psi_{\uparrow}(z)+\omega_{\downarrow}\psi^{\dag}_{\downarrow}(z)\psi_{\downarrow}(z)\bigg)+\omega_pa^{\dag}a\bigg]\bigg\},
\end{equation}
where $\delta=\omega_d-\omega_{0}$ is the detuning of the drive to the spin-$1/2$ transition.
After carrying out the unitary transformation described above, the Hamiltonian takes the following form,
\begin{equation}
    \begin{aligned}
    H&\approx -\int\mathrm{d}z(\hbar\Delta_{\downarrow})\psi^{\dag}_e(z)\psi_e(z)+\sum_{\sigma=\uparrow,\downarrow,e}\int\mathrm{d}z\psi^{\dag}_{\sigma}(z)\bigg[\frac{p^2}{2M}+V_0\sin^2(k_lz)+Mgz\bigg]\psi_{\sigma}(z)\\
    &+\int dz\bigg[\hbar\mathcal{G}_{\uparrow}(z)a\psi^{\dag}_e(z)\psi_{\uparrow}(z)\mathrm{e}^{-i(\omega_0+\delta)t}+\hbar\mathcal{G}_{\downarrow}(z)a\psi^{\dag}_e(z)\psi_{\downarrow}(z)+\mathrm{h.c.}\bigg]+\int dz\bigg[\frac{\hbar\Omega}{2}\psi^{\dag}_{\uparrow}(z)\psi_{\downarrow}(z)+\mathrm{h.c.}\bigg]\\
    &-\int dz(\hbar\delta)\psi^{\dag}_{\uparrow}(z)\psi_{\uparrow}(z)-\hbar\Delta_ca^{\dag}a+\hbar(\varepsilon a^{\dag}+\varepsilon^{*}a),\\
    \end{aligned}
\end{equation}
where $\Delta_c=\omega_p-\omega_c$ is the cavity detuning, $\Delta_{\uparrow,\downarrow}=\omega_p-\omega_e+\omega_{\uparrow,\downarrow}$ are the detunings of the pump from the corresponding electronic excited state to ground transitions. We assume $\hbar \Delta_{\uparrow,\downarrow}$ are much larger than the motional energy of the atoms. 
Here we focus on the case that $\Delta_{\uparrow,\downarrow}\gg \mathcal{G}^0_{\uparrow,\downarrow}\sqrt{\langle a^{\dag}a\rangle}$ and $\Delta_{\uparrow,\downarrow}\gg \gamma$, which means the excited state population and the decoherence induced by atomic spontaneous emission are negligible. 
In this limit we can adiabatically eliminate the atomic excited state $|e\rangle$ by averaging out the largest frequency scale $\Delta_{\uparrow,\downarrow}$ and $\omega_0$ in this system (see Ref.~\cite{james2007}). This leads to the following effective Hamiltonian acting in the ground state manifold \cite{Maschler2005,Prasanna2009},
\begin{equation}
    \begin{aligned}
    H&=\sum_{\sigma=\uparrow,\downarrow}\int dz\psi^{\dag}_{\sigma}(z)\left[\frac{p^2}{2M}+V_0\sin^2(k_lz)+Mgz+\frac{\hbar|\mathcal{G}_{\sigma}(z)|^2}{\Delta_{\sigma}}a^{\dag}a\right]\psi_{\sigma}(z)\\
    &+\int dz\bigg[\frac{\hbar\Omega}{2}\psi^{\dag}_{\uparrow}(z)\psi_{\downarrow}(z)+\mathrm{h.c.}\bigg]-\int dz(\hbar\delta)\psi^{\dag}_{\uparrow}(z)\psi_{\uparrow}(z)-\hbar\Delta_ca^{\dag}a+\hbar(\varepsilon a^{\dag}+\varepsilon^{*}a).\\
    \end{aligned}
\end{equation}
It's worth to mention that for simplicity we take only one atomic excited state into account in this derivation. For practical experimental system, one might need to sum over the contributions from all relevant excited states.

Assuming all the atoms are located in the ground band of the lattice, in tight-binding limit the single-particle eigenstates of the system can be described by Wannier-Stark states $|\phi_n\rangle\,(n\in\mathbb{Z})$ \cite{gluck2002}:
\begin{equation}
    E_n=Mga_ln, \quad \phi_n(z)=\sum_m\mathcal{J}_{m-n}\bigg(\frac{2J_0}{Mga_l}\bigg)w(z-ma_l).
\end{equation}
Here, $\mathcal{J}_n(x)$ is the Bessel function of the first kind, $J_0\approx (4/\sqrt{\pi})E_R^{1/4}V_0^{3/4}\exp[-2\sqrt{V_0/E_R}]$ is the nearest-neighbor tunneling couplings, $a_l=\lambda_l/2$ is the lattice spacing, and $w(x)$ is the ground band Wannier function. 
Here $E_R=\hbar^2k_l^2/2M$ is the atomic recoil energy.  
Then we expand the atomic field operators $\psi_{\sigma}(z)$ in terms of Wannier-Stark states,
\begin{equation}
    \psi_{\sigma}(z)=\sum_nc_{n\sigma}\phi_n(z),
\end{equation}
where $c_{n\sigma}$ annihilates an atom of spin $\sigma$ in the state $|\phi_n\rangle$. 
We consider the case that $Mga_l\gg \hbar|\mathcal{G}_{\uparrow,\downarrow}^{0}|^2\langle a^{\dag}a\rangle/\Delta_{\uparrow,\downarrow}$, which means cavity-induced AC Stark shifts in atomic ground state are smaller than $Mga_l$, the energy difference between adjacent Wannier-Stark ladder states. This implies all the atoms are frozen in their initial Wannier-Stark states (see section S4 for validity of this approximation), and we only need to consider the dynamics in the carrier transition $|\downarrow;\phi_n\rangle\leftrightarrow|\uparrow;\phi_{n}\rangle$.
Turning off the Rabi drive $\Omega$ after initial state preparation, the Hamiltonian can be simplified into the following form,
\begin{equation}
    \tilde{H}=-\hbar\tilde{\Delta}_ca^{\dag}a+\hbar(\varepsilon a^{\dag}+\varepsilon^{*}a)+\sum_n\bigg(-\hbar\delta+\hbar\eta_na^{\dag}a\bigg)S^z_n,
\end{equation}
and 
\begin{equation}
    \begin{gathered}
        \tilde{\Delta}_c=\Delta_c-\sum_n\frac{N_n}{2}\bigg(\frac{|\mathcal{G}_{\uparrow}^0|^2}{\Delta_{\uparrow}}+\frac{|\mathcal{G}_{\downarrow}^0|^2}{\Delta_{\downarrow}}\bigg)\int dz\cos^2(k_cz)\phi_{n}(z)\phi_{n}(z),\\
        \eta_n=\bigg(\frac{|\mathcal{G}_{\uparrow}^0|^2}{\Delta_{\uparrow}}-\frac{|\mathcal{G}_{\downarrow}^0|^2}{\Delta_{\downarrow}}\bigg)\int dz\cos^2(k_cz)\phi_{n}(z)\phi_{n}(z).\\
    \end{gathered}
\end{equation}
And the spin operators are defined as follows,
\begin{equation}
    \begin{gathered}
        S_n^x=\frac{1}{2}(c_{n\uparrow}^{\dag}c_{n\downarrow}+c_{n\downarrow}^{\dag}c_{n\uparrow}), \quad S_n^y=-\frac{i}{2}(c_{n \uparrow}^{\dag}c_{n\downarrow}-c_{n\downarrow}^{\dag}c_{n\uparrow}),\\
        S_n^z=\frac{1}{2}(c_{n\uparrow}^{\dag}c_{n\uparrow}-c_{n\downarrow}^{\dag}c_{n\downarrow}), \quad N_n=c_{n\uparrow}^{\dag}c_{n\uparrow}+c_{n\downarrow}^{\dag}c_{n\downarrow}.\\
    \end{gathered}
\end{equation}

Assuming the ground band Wannier function $w(x)$ are localized on a single lattice site, the integral in $\eta_n$ can be evaluated analytically,
\begin{equation}
    \int dz\cos^2(k_cz)\phi_n(z)\phi_m(z)\approx \frac{1}{2}\bigg[\delta_{nm}+\mathcal{C}\mathcal{J}_{n-m}\bigg(\frac{4J_0}{Mga_l}\sin(\varphi/2)\bigg)\cos\bigg(\frac{(n+m)\varphi}{2}+\frac{(n-m)\pi}{2}\bigg)\bigg],
\end{equation}
where $\varphi=2k_ca_l$, and $\mathcal{C}=\int dz\,\mathrm{e}^{2ik_cz}|w(z)|^2$. Thus the analytic expression for $\eta_n$ is 
\begin{equation}
    \eta_n=\frac{1}{2}\bigg[\frac{|\mathcal{G}_{\uparrow}^0|^2}{\Delta_{\uparrow}}-\frac{|\mathcal{G}_{\downarrow}^0|^2}{\Delta_{\downarrow}}\bigg]\bigg[1+\mathcal{C}\mathcal{J}_{0}\bigg(\frac{4J_0}{Mga_l}\sin(\varphi/2)\bigg)\cos(n\varphi)\bigg].
\end{equation}
which leads to the magic lattice condition to achieve uniform $\eta_n$,
\begin{equation}
    \mathcal{J}_{0}\bigg(\frac{4J_0}{Mga_l}\sin(\varphi/2)\bigg)=0.
\end{equation}
Under this magic lattice condition, we can replace $\eta_n$ by $\eta$, which is given by
\begin{equation}
    \eta=\frac{1}{2}\bigg[\frac{|\mathcal{G}_{\uparrow}^0|^2}{\Delta_{\uparrow}}-\frac{|\mathcal{G}_{\downarrow}^0|^2}{\Delta_{\downarrow}}\bigg].
\end{equation}

\section{Adiabatic elimination of cavity mode}
We consider the cavity mode has decay rate $\kappa$, so the dynamics of this system can be described by the following master equation, 
\begin{equation}
    \begin{gathered}
        \frac{d}{dt}\rho=-\frac{i}{\hbar}[\tilde{H},\rho]+L\rho L^{\dag}-\frac{1}{2}\{L^{\dag}L,\rho\},\\
        \tilde{H}/\hbar=-\tilde{\Delta}_ca^{\dag}a+\varepsilon a^{\dag}+\varepsilon^{*}a+\sum_n\bigg(-\delta+\eta_na^{\dag}a\bigg)S^z_n,\\
        L=\sqrt{\kappa}\,a.\\
    \end{gathered} 
\end{equation}
If the cavity detuning $\tilde{\Delta}_c$ is now the largest frequency scale, we can proceed to adiabatically eliminate the cavity mode. 
For that first we expand the cavity field operator $a$ as a sum of its steady state value $\alpha=\langle a\rangle$ and quantum fluctuation $b$,
\begin{equation}
    a=\alpha+b, \quad \alpha=\frac{\varepsilon}{\tilde{\Delta}_c+i\kappa/2}.
\end{equation}
Assuming $\tilde{\Delta}_c\gg \sum_n\eta_n \langle S^z_n\rangle$, the Hamiltonian and jump operators become
\begin{equation}
    \begin{gathered}
        \tilde{H}/\hbar\approx-\tilde{\Delta}_cb^{\dag}b+\sum_n\bigg(-\delta+\eta_n|\alpha|^2+\eta_n\alpha^{*}b+\eta_nb^{\dag}\alpha\bigg)S^z_n,\\
        L=\sqrt{\kappa}\,b.\\
    \end{gathered}
\end{equation}
Now we focus on the initial condition where the collective spin is aligned along the equator, and the fluctuation of $S^z$ is given by $\sqrt{N}/2$. 
In this case if $\tilde{\Delta}_c\gg \eta\alpha\sqrt{N}$ then $\langle b\rangle\ll 1$, and therefore we can adiabatically eliminate the photon excitation (see Ref.~\cite{reiter2012}) and obtain an effective master equation in zero-photon subspace,
\begin{equation}
    \begin{gathered}
        \frac{d}{dt}\rho=-\frac{i}{\hbar}[H_{\mathrm{eff}},\rho]+L_{\mathrm{eff}}\rho L_{\mathrm{eff}}^{\dag}-\frac{1}{2}\{L_{\mathrm{eff}}^{\dag}L_{\mathrm{eff}},\rho\},\\
        H_{\mathrm{eff}}/\hbar=-\sum_n(\delta_k-\eta_n|\alpha|^2)S_n^z+\sum_{nm}\frac{\eta_n\eta_m|\alpha|^2\tilde{\Delta}_c}{\tilde{\Delta}_c^2+\kappa^2/4}S^z_nS^z_m,\\
        L_{\mathrm{eff}}=\sqrt{\kappa}\sum_n\frac{\eta_n\alpha}{\tilde{\Delta}_c+i\kappa/2}S_n^z.\\
    \end{gathered}
    \label{eq:eff}
\end{equation}

\section{Effects of thermal distribution on magic lattice condition}
In the main text, we discuss the magic lattice condition to achieve uniform one-axis twisting,
\begin{equation}
    \mathcal{J}_0\bigg(\frac{4J_0}{Mga_l}\sin(\varphi/2)\bigg)=0,
    \label{magic}
\end{equation}
assuming a separable confinement potential and tunneling only along the gravity direction. 
For practical experimental system, the Gaussian geometry of the laser beams inevitably couple the vertical and radial wave-functions. In this case therefore we have to consider the thermal distribution of atoms in the transverse direction, which leads to an imperfect cancellation of the inhomogeneity. 
The Gaussian profile of the lattice and cavity beams leads to the following imperfections:
\begin{itemize}
    \item Atoms in different radial modes have a slightly different nearest-neighbor tunneling rate.
    \item Atoms in different radial modes feel a  differential AC-Stark shifts and thus extra  inhomogeneities. 
\end{itemize}

First we focus on the Gaussian beam profile of a 1D lattice, which leads to the following trapping potential,
\begin{equation}
    V_0(\mathbf{r},z)=\left\{\begin{array}{cc}
    V_0-V_0\cos^2(k_lz)\exp(-2r^2/w_l^2) & (\text{red-detuned lattice})\\
    V_0\sin^2(k_lz)\exp(-2r^2/w_l^2) & (\text{blue-detuned lattice})\\
    \end{array}\right.,
\end{equation}
where $k_l=2\pi/\lambda_l$ is the lattice wave number, $w_l$ is the beam waist, and $V_0>0$ is the lattice depth. Using a technique similar to the one used in Ref.~\cite{cox2016spatially}, one can also introduce an additional radial trapping potential $V_r(\mathbf{r},z)=M\omega_{r1}^2r^2/2$ from the optical cavity. 
Expanding the total trapping potential $V(\mathbf{r},z)=V_0(\mathbf{r},z)+V_r(\mathbf{r},z)$ to second order of $r$, we have
\begin{equation}
    V(\mathbf{r},z)\approx V_0\sin^2(k_lz)+\frac{1}{2}M\omega_r^2r^2-\frac{\omega_{r0}^2}{\omega_r^2}\cdot\frac{1}{2}M\omega_r^2r^2\sin^2(k_lz).
\end{equation}
Here, the total radial trapping frequency is given by $\omega_r=\sqrt{\omega_{r0}^2+\omega_{r1}^2}$ for red-detuned lattice, and $\omega_r=\omega_{r1}$ for blue-detuned lattice. Here  $\omega_{r0}=\sqrt{4V_0/Mw_l^2}$. 

\begin{figure}[t]
    \centering
    \includegraphics[width=16cm]{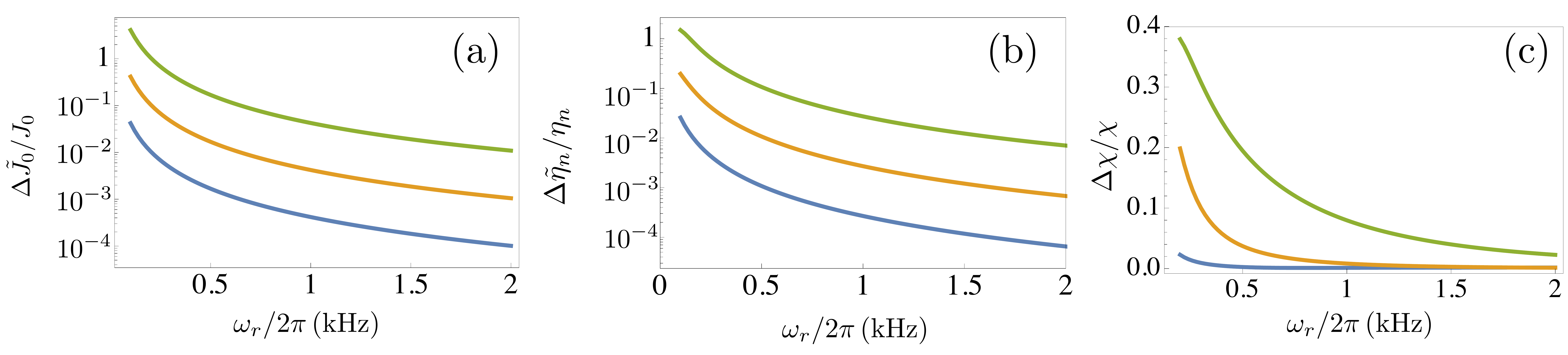}
\caption{\label{fig:RadialCorrection}The standard deviations  of (a) the tunneling rates, (b) the AC-Stark shifts and (c) the interaction strength as a function of the total radial trapping potential $\omega_{r}$ for fixed $T=0.1\mu$K (blue curves), $T=1.0\mu$K (yellow curves) and $T=10\mu$K (green curves). Here we assume the beam waists are $w_c\approx w_l=50\mu$m and the trapping depth is $V_0=6E_r$. We find the corrections can be suppressed by increasing the radial trapping frequency or lowering the temperature.}
\end{figure}

In the expression above, the first term describes the lattice potential along the axial direction. The corresponding axial eigenfunctions are Bloch functions $\phi_q(z)$ in the ground band with a band structure $E(q)/E_R=f(q/\hbar k_l,V_0/4E_R)+V_0/2E_R$, where $q\in[-\hbar k_l,\hbar k_l]$ is the quasimomentum, and the function $f$ is the characteristic Mathieu value of type A for $q\in(-\hbar k_l,\hbar k_l)$, and the characteristic Mathieu value of type B for $q=\pm \hbar k_l$. 
The second term describes the radial trapping potential as a harmonic oscillator with  eigenfunctions  $\phi_{n_x,n_y}(\mathbf{r})=\phi_{n_x}(x)\phi_{n_y}(y)$ and eigenenergies $E_{n_x,n_y}=\hbar\omega_r(n_x+n_y+1)/2$. 
The third term is the anharmonicity in the Gaussian beam profile, which couples the axial and radial degrees of freedom. 
Similar to Ref.~\cite{kolkowitz2017}, we use  first-order perturbation theory to calculate the energy corrections from the anharmonicity,
\begin{equation}
    E_{n_x,n_y}(q)=E_{n_x,n_y}+E(q)-\frac{1}{2}\frac{\omega_{r0}^2}{\omega_r^2}E_{n_x,n_y}\langle q|\sin^2(k_lz)|q\rangle.
\end{equation}
Based on Feynman–Hellman theorem, we have
\begin{equation}
    \langle q|\sin^2(k_lz)|q\rangle=\frac{1}{2}+\frac{\partial }{\partial v_0}f(\tilde{q},v_0/4),
\end{equation}
in which we define $\tilde{q}=q/\hbar k_l$, and $v_0=V_0/E_R$. Notice that in the tight-binding limit, we can calculate the nearest-neighbor tunneling by $J_0=[E(q=\pm\hbar k_l)-E(q=0)]/4$, so the correction of $J_0$ is given by
\begin{equation}
    \tilde{J}_0(n_x,n_y)=J_0+\frac{1}{8}\frac{\omega_{r0}^2}{\omega_r^2}E_{n_x,n_y}\bigg[\frac{\partial }{\partial v_0}f(\tilde{q}=0,v_0/4)-\frac{\partial }{\partial v_0}f(\tilde{q}=\pm 1,v_0/4)\bigg],
\end{equation}
where $J_0\approx (4/\sqrt{\pi})E_R^{1/4}V_0^{3/4}\exp[-2\sqrt{V_0/E_R}]$. As depicted in Fig.~\ref{fig:RadialCorrection}(a), one can suppress the correction of $J_0$ by increasing the total radial trapping frequency $\omega_r$, or lowering the temperature.

With respect to the cavity mode we used to mediate one-axis twisting interaction, and this cavity mode has the finite beam waist $w_c$. Notice that since the 1D lattice is also injected into the cavity, it's reasonable to assume $w_c\approx w_l$. In this case  we can estimate the differential AC-Stark shifts by
\begin{equation}
    \tilde{\eta}_n(n_x,n_y)=\eta_n\int dxdy\exp\bigg(-\frac{2r^2}{w_c^2}\bigg)[\phi_{n_x}(x)\phi_{n_y}(y)]^2,
\end{equation}
where
\begin{equation}
    \phi_m(x)=\frac{1}{\sqrt{2^mm!}}\bigg(\frac{M\omega_r}{\pi\hbar}\bigg)^{1/4}\mathrm{e}^{-M\omega_r x^2/2\hbar}H_m\bigg(\sqrt{\frac{M\omega_r}{\hbar}}x\bigg).
\end{equation}

Using the following identity of Hermite polynomials,
\begin{equation}
    \int dx\,\mathrm{e}^{-2\alpha^2 x^2}[H_m(x)]^2=2^{m-1/2}\alpha^{-(2m+1)}(1-2\alpha^2)^{m}\Gamma\bigg(m+\frac{1}{2}\bigg){}_2F_1\bigg(-m,-m;\frac{1}{2}-m;\frac{\alpha^2}{2\alpha^2-1}\bigg),
\end{equation}
where ${}_2F_1(a,b;c;x)$ is hypergeometric function, we get
\begin{equation}
    \tilde{\eta}_n(n_x,n_y)=\eta_n h(n_x)h(n_y),
\end{equation}
where
\begin{equation}
    h(m)=\frac{1}{\sqrt{2\pi}}\frac{1}{m!}\alpha^{-(2m+1)}(1-2\alpha^2)^{m}\Gamma\bigg(m+\frac{1}{2}\bigg){}_2F_1\bigg(-m,-m;\frac{1}{2}-m;\frac{\alpha^2}{2\alpha^2-1}\bigg),
\end{equation}
\begin{equation}
    \alpha=\sqrt{\frac{1}{2}+\frac{\hbar}{M\omega_r w_c^2}}.
\end{equation}
As depicted in Fig.~\ref{fig:RadialCorrection}(b), one can suppress the the differential AC-Stark shifts by increasing the total radial trapping frequency $\omega_r$, or lowering the temperature. Finally, we calculate the inhomogeneties in the OAT interaction strength [Fig.~\ref{fig:RadialCorrection}(c)] combining the corrections in tunneling rates and differential AC Stark shifts.

\section{Validity of effective spin models}
\subsection{Tight-binding approximation}
Our protocol in the main text works in relatively shallow lattices. However, all the discussions above were based on the tight-binding approximation which neglects the next nearest neighbour and higher order  tunneling processes. It's therefore important to check the validity of the tight-binding approximation for typical lattice depths in our protocol. We denote the tunneling rate between site $n$ and site $n+m$ in the ground band as $J_{0,m}/\hbar$, which can be calculated by 
\begin{equation}
    \frac{J_{0,m}}{E_R}=-\frac{1}{2}\int_{-1}^{1}d\tilde{q}\,\mathrm{e}^{-im\pi\tilde{q}}f(\tilde{q},v_0/4).
\end{equation}
in which the function $f$ related to characteristic Mathieu value is defined in the previous section. For $^{87}$Rb atoms and $532$nm lattice, we have $|J_{0,2}|/|J_{0,1}|=0.038$ at $V_0=6.0E_R$, and $|J_{0,2}|/|J_{0,1}|=0.105$ at $V_0=2.9E_R$. This result shows that the nearest neighbor tunneling rate is dominant for the typical lattice depths in our discussion, which agrees with the key idea of tight-binding approximation. And we can also numerically check that the cancellation of inhomogeneities under magic lattice condition is still approximately valid even if we include all the possible tunneling processes [see Fig.~\ref{fig:Validity}(a,b)].  

\subsection{Hopping between Wannier-Stark states}
In the main text, we consider an approximation that all the atoms are frozen in the Wannier-Stark (WS) states from Eq.~(1) to Eq.~(2), assuming the cavity-induced AC Stark shifts are smaller than the energy gap $Mga_l$ between WS states. Here we provide the justifications of this approximation in detail. We denote the corrections to Eq.~(2) in the main text by $H_{\mathrm{corr}}$, which takes the following form,
\begin{equation}
    H_{\mathrm{corr}}=\sum_{n\sigma}(Mga_ln)c_{n\sigma}^{\dag}c_{n\sigma}+\sum_{m\neq n,\sigma}K^{nm}_{\sigma}a^{\dag}ac_{n\sigma}^{\dag}c_{m\sigma},
\end{equation}
where $K^{nm}_{\sigma}$ with $m\neq n$ is given by
\begin{equation}
    K^{nm}_{\sigma}=\frac{1}{2}\frac{\hbar|\mathcal{G}^0|^2}{\Delta_{\sigma}}\mathcal{C}\mathcal{J}_{n-m}\bigg(\frac{4J_0}{Mga_l}\sin(\varphi/2)\bigg)\cos\bigg(\frac{(n+m)\varphi}{2}+\frac{(n-m)\pi}{2}\bigg).
\end{equation}

If we replace the cavity field operator $a$ by its steady state value $\alpha$, $H_{\mathrm{corr}}$ takes the similar form as the single-particle Hamiltonian in lattice oriented in vertical direction, but here we are considering hopping between WS states instead of Wannier states. And our approximation is mainly based on the fact that hopping between WS states is highly suppressed by the large energy gap $Mga_l$. To estimate the suppression, we would like to neglect the terms beyond nearest neighbor hopping, as these terms will eventually be suppressed by the Bessel function $\mathcal{J}_{n-m}$, and then we also replace the inhomogeneous $K^{nm}_{\sigma}$ by its peak value $K$. If we initialize an atom at the WS state $|\phi_{n=0}\rangle$, the probability in find this atom at WS state $|\phi_{n}\rangle$ is given by
\begin{equation}
    P_n(t)=\mathcal{J}_n\bigg[\frac{4K|\alpha|^2}{Mga_l}\bigg|\sin\bigg(\frac{Mga_l t}{2\hbar}\bigg)\bigg|\bigg]^2.
\end{equation}
Average over the fast frequency scale $Mga_l/\hbar$, we can estimate the probability in WS state $|\phi_{n=1}\rangle$ by $\overline{P}_1\approx 2K^2|\alpha|^4/(Mga_l)^2$. For $^{87}$Rb atoms and $532$nm lattice, if we choose the magic lattice condition $V_0=6.0E_R$, and set the cavity-induced AC Stark shift $\eta|\alpha|^2=0.2 Mga_l/\hbar$ (assuming $\mathcal{G}^0_{\uparrow}=\mathcal{G}^0_{\downarrow}$ and $\Delta_{\uparrow}=-\Delta_{\downarrow}$ to estimate $K$), we have $\overline{P}_1\approx 0.5\%$. And this is why we can assume all the atoms are frozen in the initial WS states.

\begin{figure}[t]
    \centering
    \includegraphics[width=17cm]{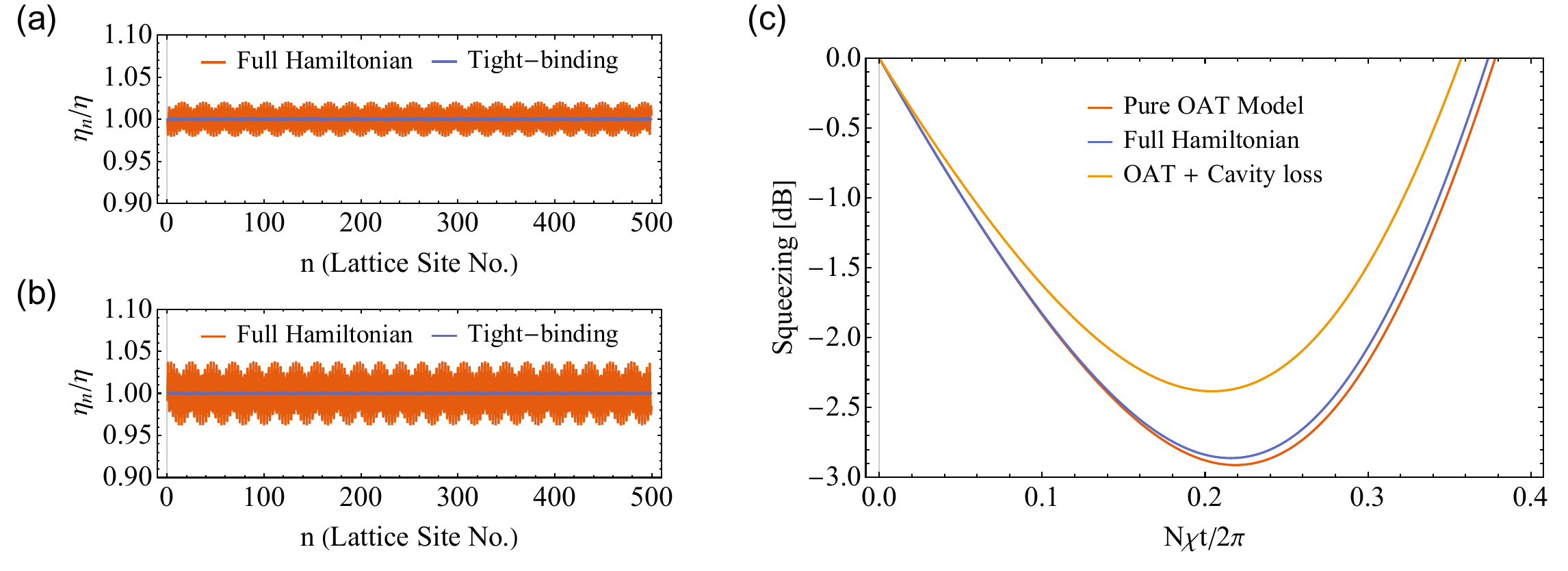}
\caption{\label{fig:Validity} (a,b) Beyond tight-binding model effects on the dispersive atom-light coupling $\eta_n$ for lattice depth (a) $V_0=6.0E_R$ and (b) $V_0=2.9E_R$. (c) Modifications on the achievable spin squeezing by cavity induced WS couplings for $4$ particles in a 4-site lattice. The orange curve shows the spin squeezing generated by pure one-axis twisting (OAT) model, the blue curve includes the effect of correction terms $\tilde{H}_{\mathrm{corr}}$ with $\eta|\alpha|^2=0.2 Mga_l/\hbar$, while the yellow curve includes the effect of cavity loss with $\Gamma_z/\chi=0.1$.}
\end{figure}

One can also step further to analyze the quantum fluctuations of cavity field $a$ in $H_{\mathrm{corr}}$. If we add $H_{\mathrm{corr}}$ back to Eq.~(2) in the main text, and then adiabatically eliminate the cavity field based on section S2, the corrections to Eq.~(3) in the main text would be
\begin{equation}
    \tilde{H}_{\mathrm{corr}}=\sum_{n\sigma}(Mga_ln)c_{n\sigma}^{\dag}c_{n\sigma}+\sum_{m\neq n,\sigma}K^{nm}_{\sigma}|\alpha|^2c_{n\sigma}^{\dag}c_{m\sigma}+\sum_{nmpq}'\sum_{\sigma\sigma'}\frac{K^{nm}_{\sigma}K^{pq}_{\sigma'}|\alpha|^2}{\tilde{\Delta}_c}(c_{n\sigma}^{\dag}c_{m\sigma})(c_{p\sigma'}^{\dag}c_{q\sigma'}),
\end{equation}
where $\sum_{nmpq}'$ means neglecting the terms with $n=m$ and $p=q$. Similar to the discussions above, the terms violate the energy conservation are highly suppressed by the large energy gap $Mga_l$. And one may only need to consider the effect of some resonant terms like $(c_{n+1,\sigma}^{\dag}c_{n\sigma})(c_{p-1,\sigma'}^{\dag}c_{p\sigma'})$, which play a role similar to dephasing because there are no spin flip processes in these correction terms. As depicted in Fig.~\ref{fig:Validity}(c), we compare the spin squeezing achieved in the ideal one-axis twisting model (see Eq.~(3) in the main text) with exact diagonalization of the full Hamiltonian (add back $\tilde{H}_{\mathrm{corr}}$) for $4$ particles in a 4-site lattice with magic lattice condition. The effect of all these corrections turns out to be negligible for the spin squeezing generation, as it is much smaller than the effect of collective dephasing with $\Gamma_z/\chi=0.1$ due to cavity loss (see section S5).

\section{Experimental considerations}
\subsection{Cavity loss}
In section S2, we already include cavity loss in the derivation of Eq.~(\ref{eq:eff}). Now we discuss the effect of cavity loss in our scheme for  quantum-enhanced gravimetry. For simplicity, we focus on pure OAT interaction with homogeneous couplings, which can be achieved with the magic lattice condition and spin echo pulses. In this case the effective master equation takes the following form,
\begin{equation}
    \frac{\mathrm{d}}{\mathrm{d}t}\rho=-i[\chi S^zS^z,\rho]+\Gamma_z\bigg[S^z\rho S^z-\frac{1}{2}\{S^zS^z,\rho\}\bigg],
    \label{eq:cavityloss}
\end{equation}
where
\begin{equation}
    \chi=\frac{\eta^2|\alpha|^2\tilde{\Delta}_c}{\tilde{\Delta}_c^2+\kappa^2/4}, \quad \Gamma_z=\frac{\kappa}{\tilde{\Delta}_c}\chi=\frac{\eta^2|\alpha|^2\kappa}{\tilde{\Delta}_c^2+\kappa^2/4}.
    \label{eqp}
\end{equation}
Similar to Ref.~\cite{Ji2013}, it is convenient to expand the density matrix in the collective spin basis $|S,m\rangle\langle S,n|$,
\begin{equation}
    \rho=\sum_{mn}\rho_{mn}|S,m\rangle\langle S,n|,
\end{equation}
and Eq.~(\ref{eq:cavityloss}) becomes
\begin{equation}
    \frac{\mathrm{d}}{\mathrm{d}t}\rho_{mn}=\bigg[-i\chi(m^2-n^2)-\frac{\Gamma_z}{2}(m-n)^2\bigg]\rho_{mn}.
\end{equation}

First, we calculate the quantum noise for phase measurement $\sigma_p=(\Delta S^y/S)_{\phi\rightarrow 0}$. Starting from the initial state $|S^x=N/2\rangle$, we let the system evolve on OAT interaction for time $t_0$, then accumulate a phase in the rotation about $\hat{y}$ axis, $\tilde{R}_y^{\phi}=\mathrm{e}^{-i\phi S^y}$, and finally perform time reversal on the OAT interaction part. The twisting echo above leads to the following density matrix,
\begin{equation}
    \rho_{mn}(2t_0)|_{\phi\rightarrow 0}=\rho_{mn}(0)\exp\Big[-(m-n)^2\Gamma_zt_0\Big],
\end{equation}
where $\rho_{mn}(0)=c_mc_n^{*}$ with
\begin{equation}
    c_m=\frac{1}{2^S}\sqrt{\frac{(2S)!}{(S+m)!(S-m)!}}.
\end{equation}
So we have
\begin{equation}
    \begin{aligned}
    (\Delta S^y)^2_{\phi\rightarrow 0}&=\mathrm{Tr}[S^yS^y\rho_{\phi\rightarrow 0}]=\frac{1}{2}\bigg[S\bigg(S+\frac{1}{2}\bigg)-S\bigg(S-\frac{1}{2}\bigg)\mathrm{e}^{-4\Gamma_zt_0}\bigg]\\
    &\approx \frac{S}{2}\Big(1+4S\Gamma_zt_0\Big),\\
    \end{aligned}
    \label{eq:loss1}
\end{equation}
in which the approximation is valid if $\Gamma_z t_0\ll 1$ and $S\gg 1$. Here we also used the fact that $\langle S^y\rangle_{\phi\rightarrow 0}=0$.

Then, we calculate the amplification factor $G=(\partial_{\phi}\langle S^y\rangle/S)_{\phi\rightarrow 0}$. It's convenient to rewrite Eq.~(\ref{eq:cavityloss}) using the Lindblad superoperator $\mathcal{L}_{\chi}$, $\partial_t\rho=\mathcal{L}_{\chi}(\rho)$. Notice that $\langle S^y\rangle_{\phi\rightarrow 0}=0$, for small $\phi$ we have 
\begin{equation}
    \begin{aligned}
    \langle S^y\rangle&=\mathrm{Tr}\bigg[S^y\mathrm{e}^{\mathcal{L}_{-\chi}t_0}\Big(\mathrm{e}^{-i\phi S^y}\mathrm{e}^{\mathcal{L}_{\chi}t_0}(\rho)\mathrm{e}^{i\phi S^y}\Big)\bigg]\\
    &\approx i\phi\bigg\{\mathrm{Tr}\bigg[S^y\mathrm{e}^{\mathcal{L}_{-\chi}t_0}\Big(\mathrm{e}^{\mathcal{L}_{\chi}t_0}(\rho)S^y\Big)\bigg]-\mathrm{Tr}\bigg[S^y\mathrm{e}^{\mathcal{L}_{-\chi}t_0}\Big(S^y\mathrm{e}^{\mathcal{L}_{\chi}t_0}(\rho)\Big)\bigg]\bigg\},\\
    \end{aligned}
\end{equation}
This yields
\begin{equation}
    (\partial_{\phi}\langle S^y\rangle)_{\phi\rightarrow 0}=S(2S-1)\sin(\chi t_0)\cos^{2S-2}(\chi t_0)\cosh(\Gamma_zt_0)\mathrm{e}^{-3\Gamma_z t_0/2}.
    \label{eq:loss2}
\end{equation}
The metrological gain $\xi^{-2}=1/[N(\Delta\phi)^2]$ can be calculated based on Eq.~(\ref{eq:loss1}) and Eq.~(\ref{eq:loss2}) as: 
\begin{equation}
    \xi^2\approx \frac{1+2N\Gamma_z t_0}{(N\chi t_0)^2}+\frac{1}{N}+\frac{1}{2}(\chi t_0)^2.
    \label{eq:cavitygain}
\end{equation}

\subsection{Spontaneous emission}
Here we discuss the effect of spontaneous emission on our gravimetry protocol focusing again on the pure OAT model with homogeneous couplings, which leads single-particle decoherence in contrast to the collective dephasing described in the previous subsection.
Single-particle decoherence mechanisms for atoms off-resoantly coupled to excited states include spin flip or dephasing processes emerging from Raman or Rayleigh scattering into free space respectively, which effectively transfer the atomic spin states outside the collective Dicke manifold.  
These processes can be described by the following master equation acting on the spin degrees of freedom,
\begin{equation}
    \begin{aligned}
    \frac{\mathrm{d}}{\mathrm{d}t}\rho&=-i\Big[\frac{\chi}{2}\sum_{j<k}\sigma^z_j\sigma^z_k,\rho\Big]+\gamma_{+}\sum_j\bigg[\sigma_j^{+}\rho\sigma_j^{-}-\frac{1}{2}\{\sigma_j^{-}\sigma_j^{+},\rho\}\bigg]\\
    &+\gamma_{-}\sum_j\bigg[\sigma_j^{-}\rho\sigma_j^{+}-\frac{1}{2}\{\sigma_j^{+}\sigma_j^{-},\rho\}\bigg]+\frac{\gamma_z}{4}\sum_j\bigg[\sigma^z_j\rho\sigma^z_j-\rho\bigg].\\
    \end{aligned}
    \label{eq:spontaneous}
\end{equation}
The analytic solution of Eq.~(\ref{eq:spontaneous}) has been reported in Ref.~\cite{FossFeig2013} by summing over all possible quantum trajectories. Here we present an alternative way by solving Heisenberg equations of motion to all order. Let's first focus on the terms that commute with the Hamiltonian, we have
\begin{equation}
    \frac{\mathrm{d}}{\mathrm{d}t}\langle\sigma^z_1\rangle=-(\gamma_{+}+\gamma_{-})\langle \sigma^z_1\rangle+(\gamma_{+}-\gamma_{-}) \quad\Rightarrow\quad \langle\sigma^z_1\rangle=\frac{\gamma_{+}-\gamma_{-}}{\gamma_{+}+\gamma_{-}}\bigg[1-\mathrm{e}^{-(\gamma_{+}+\gamma_{-})t}\bigg],
\end{equation}
\begin{equation}
    \frac{\mathrm{d}}{\mathrm{d}t}\langle\sigma^z_1\sigma^z_2\rangle=-2(\gamma_{+}+\gamma_{-})\langle\sigma^z_1\sigma^z_2\rangle+2(\gamma_{+}-\gamma_{-})\langle\sigma^z_1\rangle \quad\Rightarrow\quad \langle\sigma^z_1\sigma^z_2\rangle=\langle\sigma^z_1\rangle^2,
\end{equation}
\begin{equation}
    \frac{\mathrm{d}}{\mathrm{d}t}\langle\sigma^{+}_1\sigma^{-}_2\rangle=-2\Gamma\langle\sigma^{+}_1\sigma^{-}_2\rangle \quad\Rightarrow\quad \langle\sigma^{+}_1\sigma^{-}_2\rangle=\frac{1}{4}\mathrm{e}^{-2\Gamma t},
    \label{eq48}
\end{equation}
where $\Gamma=(\gamma_{+}+\gamma_{-}+\gamma_z)/2$, and we consider the initial state as $|S^x=N/2\rangle$. Then we consider $\langle \sigma_1^{+}\rangle$ and the higher order correlators generated in the Heisenberg equations of motion,
\begin{equation}
    \frac{\mathrm{d}}{\mathrm{d}t}\begin{pmatrix}
        \langle\sigma_1^{+}\rangle\\
        \langle\sigma_1^{+}\sigma_2^{z}\rangle\\
        \langle\sigma_1^{+}\sigma_2^{z}\sigma_3^{z}\rangle\\
        \vdots\\
        \langle\sigma_1^{+}\sigma_2^{z}\cdots \sigma_{N}^{z}\rangle\\
    \end{pmatrix}=\begin{pmatrix}
        -\Gamma & i\chi(N-1) & 0 & 0 & \cdots & 0 & 0\\
        J & -\Gamma-\Gamma_r & i\chi(N-2) & 0 & \cdots & 0 & 0\\
        0 & 2J & -\Gamma-2\Gamma_r & i\chi(N-3) & \cdots & 0 & 0\\
        \vdots & \vdots & \vdots & \vdots & & \vdots & \vdots\\
        0 & 0 & 0 & 0 & \cdots & (N-1)J & -\Gamma-(N-1)\Gamma_r\\
    \end{pmatrix}\begin{pmatrix}
        \langle\sigma_1^{+}\rangle\\
        \langle\sigma_1^{+}\sigma_2^{z}\rangle\\
        \langle\sigma_1^{+}\sigma_2^{z}\sigma_3^{z}\rangle\\
        \vdots\\
        \langle\sigma_1^{+}\sigma_2^{z}\cdots \sigma_{N}^{z}\rangle\\
    \end{pmatrix},
\end{equation}
where we define $J=\gamma_{+}-\gamma_{-}+i\chi$, and $\Gamma_r=\gamma_{+}+\gamma_{-}$. We can solve the differential equations above via a trial solution of the form,
\begin{equation}
    \begin{gathered}
    \langle\sigma_1^{+}\rangle=\frac{1}{2}[f(t)]^{N-1}\mathrm{e}^{-\Gamma t}, \quad \langle\sigma_1^{+}\sigma_2^{z}\rangle=\frac{1}{2}[f(t)]^{N-2}\frac{f'(t)}{i\chi}\mathrm{e}^{-\Gamma t},\\
    \langle\sigma_1^{+}\sigma_2^{z}\sigma_3^{z}\rangle=\frac{1}{2}[f(t)]^{N-3}\bigg(\frac{f'(t)}{i\chi}\bigg)^2\mathrm{e}^{-\Gamma t}, \quad \cdots, \quad \langle\sigma_1^{+}\sigma_2^{z}\cdots \sigma_{N}^{z}\rangle=\frac{1}{2}\bigg(\frac{f'(t)}{i\chi}\bigg)^{N-1}\mathrm{e}^{-\Gamma t}.\\
    \end{gathered}
\end{equation}
This trial solution simplifies the coupled differential equations above into a single equation of the form,
\begin{equation}
    f''(t)+\Gamma_rf'(t)-i\chi Jf(t)=0.
    \label{eq:fpp}
\end{equation}
For an initial state as $|S^x=N/2\rangle$, the initial condition for the differential equation is $f(0)=1$, $f'(0)=0$. A similar analysis can also be applied  to $\langle \sigma_1^{+}\sigma_2^{+}\rangle$ as follows,
\begin{equation}
    \frac{\mathrm{d}}{\mathrm{d}t}\begin{pmatrix}
        \langle\sigma_1^{+}\sigma_2^{+}\rangle\\
        \langle\sigma_1^{+}\sigma_2^{+}\sigma_3^{z}\rangle\\
        \vdots\\
        \langle\sigma_1^{+}\sigma_2^{+}\sigma_3^{z}\cdots \sigma_{N}^{z}\rangle\\
    \end{pmatrix}=\begin{pmatrix}
        -2\Gamma & 2i\chi(N-2) & 0 & 0 & \cdots & 0 & 0\\
        J' & -\Gamma-\Gamma_r & 2i\chi(N-3) & 0 & \cdots & 0 & 0\\
        \vdots & \vdots & \vdots & \vdots & & \vdots & \vdots\\
        0 & 0 & 0 & 0 & \cdots & (N-2)J' & -\Gamma-(N-2)\Gamma_r\\
    \end{pmatrix}\begin{pmatrix}
        \langle\sigma_1^{+}\sigma_2^{+}\rangle\\
        \langle\sigma_1^{+}\sigma_2^{+}\sigma_3^{z}\rangle\\
        \vdots\\
        \langle\sigma_1^{+}\sigma_2^{+}\sigma_3^{z}\cdots \sigma_{N}^{z}\rangle\\
    \end{pmatrix},
\end{equation}
where we define $J'=\gamma_{+}-\gamma_{-}+2i\chi$. Using the following trial solution,
\begin{equation}
    \begin{gathered}
    \langle\sigma_1^{+}\sigma_2^{+}\rangle=\frac{1}{4}[g(t)]^{N-2}\mathrm{e}^{-2\Gamma t}, \quad \langle\sigma_1^{+}\sigma_2^{+}\sigma_3^{z}\rangle=\frac{1}{4}[g(t)]^{N-3}\frac{g'(t)}{2i\chi}\mathrm{e}^{-2\Gamma t}\\
    \langle\sigma_1^{+}\sigma_2^{+}\sigma_3^{z}\sigma_4^{z}\rangle=\frac{1}{4}[g(t)]^{N-4}\bigg(\frac{g'(t)}{2i\chi}\bigg)^2\mathrm{e}^{-2\Gamma t}, \quad \cdots, \quad \langle\sigma_1^{+}\sigma_2^{+}\sigma_3^{z}\cdots \sigma_{N}^{z}\rangle=\frac{1}{4}\bigg(\frac{g'(t)}{2i\chi}\bigg)^{N-2}\mathrm{e}^{-2\Gamma t},\\
    \end{gathered}
\end{equation}
we simplify the coupled differential equations above to a single one in the following form,
\begin{equation}
    g''(t)+\Gamma_rg'(t)-2i\chi J'g(t)=0.
    \label{eq:gpp}
\end{equation}
For the initial state  $|S^x=N/2\rangle$, the initial condition for the differential equation is $g(0)=1$, $g'(0)=0$.

After introducing the analytic solutions for the master equation Eq.~(\ref{eq:spontaneous}), now we apply them to the twisting echo protocol. First, we calculate the quantum noise for phase measurement $\sigma_p=(\Delta S^y/S)_{\phi\rightarrow 0}$. Notice that $\langle S^y\rangle_{\phi\rightarrow 0}=0$, we have
\begin{equation}
    (\Delta S^y)^2_{\phi\rightarrow 0}=\langle S^yS^y\rangle_{\phi\rightarrow 0}=\frac{N}{4}+\frac{N(N-1)}{2}\mathrm{Re}\Big[\langle \sigma_1^{+}\sigma_2^{-}\rangle-\langle \sigma_1^{+}\sigma_2^{+}\rangle\Big],
\end{equation}
so we need to calculate $\langle \sigma_1^{+}\sigma_2^{-}\rangle$ and $\langle \sigma_1^{+}\sigma_2^{+}\rangle$ in the case of $\phi=0$. It's easy to observe Eq.~(\ref{eq48}) that the dynamics of $\langle \sigma_1^{+}\sigma_2^{-}\rangle$ does not depend on $\chi$. The solution is given by 
\begin{equation}
    \langle \sigma_1^{+}\sigma_2^{-}\rangle_{t=2t_0}=\frac{1}{4}\mathrm{e}^{-4\Gamma t_0}.
\end{equation}
As for $\langle \sigma_1^{+}\sigma_2^{+}\rangle$, since we focus on the case with balanced spin flip rates, $\gamma_{+}=\gamma_{-}\equiv\gamma_r$, we can rewrite Eq.~(\ref{eq:gpp}) into the following form,
\begin{equation}
    g''(t)+2\gamma_rg'(t)+4\chi^2g(t)=0,
    \label{eq:gppn}
\end{equation}
and applying the initial condition $g(0)=1$, $g'(0)=0$ leads to
\begin{equation}
    g(t_0)=\bigg[\cosh(\nu t_0)+\frac{\gamma_r}{\nu}\sinh(\nu t_0)\bigg]\mathrm{e}^{-\gamma_rt_0}, \quad g'(t_0)=\frac{\nu^2-\gamma_r^2}{\nu}\sinh(\nu t_0)\mathrm{e}^{-\gamma_rt_0},
\end{equation}
where $\nu=\gamma_r\sqrt{1-4\chi^2/\gamma_r^2}$. For the time reversal of OAT interaction ($\chi\rightarrow -\chi$), we can apply a modified version of trial solution as follows,
\begin{equation}
    \langle\sigma_1^{+}\sigma_2^{+}\rangle=\frac{1}{4}\mathrm{e}^{-2\Gamma t_0}[\tilde{g}(t)]^{N-2}\mathrm{e}^{-2\Gamma t}, \quad \langle\sigma_1^{+}\sigma_2^{+}\sigma_3^{z}\rangle=\frac{1}{4}\mathrm{e}^{-2\Gamma t_0}[\tilde{g}(t)]^{N-3}\frac{\tilde{g}'(t)}{2i(-\chi)}\mathrm{e}^{-2\Gamma t}, \quad\cdots,
    \label{eq:modi}
\end{equation}
in which $\tilde{g}(t)$ still satisfying the differential equation Eq.~(\ref{eq:gppn}), with initial condition $\tilde{g}(0)=g(t_0)$ and $\tilde{g}'(0)=-g'(t_0)$. So we get
\begin{equation}
    \tilde{g}(t_0)=\bigg[1-\frac{\gamma_r^2}{\nu^2}+\frac{\gamma_r^2}{\nu^2}\cosh(2\nu t_0)+\frac{\gamma_r}{\nu}\sinh(2\nu t_0)\bigg]\mathrm{e}^{-2\gamma_rt_0},
\end{equation}
which leads to
\begin{equation}
    \langle \sigma_1^{+}\sigma_2^{+}\rangle_{t=2t_0}=\frac{1}{4}\mathrm{e}^{-4\Gamma t_0}[\tilde{g}(t_0)]^{N-2}\approx \frac{1}{4}\mathrm{e}^{-4\Gamma t_0}\bigg[1-\frac{16}{3}(N-2)\gamma_r\chi^2t_0^3\bigg],
\end{equation}
in which the approximation is valid if $\gamma_rt_0\ll 1$. So we have
\begin{equation}
    (\Delta S^y)^2_{\phi\rightarrow 0}\approx \frac{N}{4}\bigg[1+\frac{8}{3}(N\chi)^2\gamma_rt_0^3\bigg].
    \label{eq:sp1}
\end{equation}

Then, we calculate the amplification factor $G=(\partial_{\phi}\langle S^y\rangle/S)_{\phi\rightarrow 0}$. Similarly we can rewrite Eq.~(\ref{eq:fpp}) into the following form, 
\begin{equation}
    f''(t)+2\gamma_rf'(t)+\chi^2f(t)=0.
    \label{eq:fppn}
\end{equation}
and then apply the initial condition $f(0)=1$, $f'(0)=0$ to calculate $f(t_0)$ and $f'(t_0)$. The next step is to apply the rotation about $\hat{y}$ axis, $\tilde{R}_y^{\phi}=\mathrm{e}^{-i\phi S^y}$, which leads to 
\begin{equation}
    \langle (\tilde{R}_y^{\phi})^{\dag}\sigma_1^{+}\tilde{R}_y^{\phi}\rangle=\langle\sigma^{+}_1\rangle_{t=t_0}\cos\phi,\quad \langle (\tilde{R}_y^{\phi})^{\dag}\sigma_1^{+}\sigma_2^z\tilde{R}_y^{\phi}\rangle=\langle \sigma_1^{+}\sigma_2^z\rangle_{t=t_0}\cos\phi-2\Big(\langle \sigma_1^{+}\sigma_2^{-}\rangle_{t=t_0}+\langle \sigma_1^{+}\sigma_2^{+}\rangle_{t=t_0}\Big)\cos\phi\sin\phi,
\end{equation}
where we use the fact that $\langle \sigma_1^y\rangle_{t=t_0}=\langle \sigma_1^z\rangle_{t=t_0}=\langle \sigma_1^x\sigma_2^z\rangle_{t=t_0}=\langle \sigma_1^x\sigma_2^y\rangle_{t=t_0}=0$. Then we apply a modified version of trial solution, similar to Eq.~(\ref{eq:modi}), to take account of the time reversal of OAT interaction. And we can similarly define $\tilde{f}(t)$ still satisfying Eq.~(\ref{eq:fppn}), with the following initial condition for small $\phi$ limit,
\begin{equation}
    \tilde{f}(0)\approx f(t_0), \quad \tilde{f}'(0)\approx-\Big[f'(t_0)-i\chi\phi\Big(1-(N-2)\chi^2t_0^2\Big)\mathrm{e}^{-\Gamma t_0}\Big],
\end{equation}
which leads to 
\begin{equation}
    \mathrm{Re}[\tilde{f}(t_0)]\approx 1-\frac{4}{3}\gamma_r\chi^2t_0^3, \quad \mathrm{Im}[\tilde{f}(t_0)]\approx \phi\Big(1-(N-2)\chi^2t_0^2\Big)(\chi t_0)\mathrm{e}^{-\Gamma t_0}\mathrm{e}^{-\gamma_rt_0}.
\end{equation}
So we have
\begin{equation}
    (\partial_{\phi}\langle S^y\rangle)_{\phi\rightarrow 0}\approx\frac{N(N-1)}{2}\mathrm{Re}[\tilde{f}(t_0)]^{N-2}\mathrm{Im}[\tilde{f}(t_0)]\mathrm{e}^{-2\Gamma t_0}\approx \frac{N(N-1)}{2}(\chi t_0)\mathrm{e}^{-4\gamma_rt_0}\mathrm{e}^{-3\gamma_zt_0/2}.
    \label{eq:sp2}
\end{equation}
And the metrological gain $\xi^{-2}=1/[N(\Delta\phi)^2]$ can be calculated based on Eq.~(\ref{eq:sp1}) and Eq.~(\ref{eq:sp2}),
\begin{equation}
    \xi^2\approx \frac{1+(8\gamma_r+3\gamma_z)t_0}{(N\chi t_0)^2}+\frac{8}{3}\gamma_rt_0.
    \label{eq:spgain}
\end{equation}

\subsection{Joint effects of cavity loss and spontaneous emission}
Now we discuss the joint effects of cavity loss and spontaneous emission, and we can show that optimal sensitivity of gravimetry only depends on the atom number $N$ and single atom cooperativity $C=4|\mathcal{G}^0|^2/\kappa\gamma$. Here, $\mathcal{G}^0$ relates to $\mathcal{G}^0_{\uparrow,\downarrow}$ defined in the main text up to Clebsch-Gordan coefficients, $\kappa$ is the cavity decay rate, and $\gamma$ is the spontaneous emission rate of the relevant electronic state coupled to the cavity. The dipole matrix element in $|\mathcal{G}^0|^2$ and $\gamma$ will eventually cancel in such a way that $C$ only depends on the details of the cavity. In our case, it is convenient to rewrite $C$ in terms of the parameters in the effective master equation in previous subsections. Notice that we can express $\Gamma=\gamma_r+\gamma_z/2$ defined in the previous subsection by $\Gamma\propto \gamma|\mathcal{G}^0_{\uparrow,\downarrow}|^2|\alpha|^2/\Delta_{\uparrow,\downarrow}^2$, and the photon-mediated  interaction strength $\chi,\Gamma_z$ [see Eq.~(\ref{eqp})] by $\chi\propto|\mathcal{G}^0_{\uparrow,\downarrow}|^4|\alpha|^2/\Delta_{\uparrow,\downarrow}^2\tilde{\Delta}_c$ and
$\Gamma_z\propto|\mathcal{G}^0_{\uparrow,\downarrow}|^4|\alpha|^2\kappa /\Delta_{\uparrow,\downarrow}^2\tilde{\Delta}_c^2$,
when $\tilde{\Delta}_c\gg \kappa$ as the case considered here.
In this limit we  can express the single atom cooperativity $C$ as 
\begin{equation}
    C=AC',\quad C'=\frac{\chi^2}{\Gamma_z\Gamma}.
    \label{eq:cooperativity}
\end{equation}
Here, $A$ is a  multiplicative constant set by appropriate  Clebsch-Gordan coefficients. 
Based on the discussions above, if we increase $\tilde{\Delta}_c$, the effect of cavity loss decreases, while the effect of spontaneous emission increases. So we can consider $\tilde{\Delta}_c$ as the tuning parameter for the relative strength of cavity loss and spontaneous emission. For convenience, we also define $d\equiv\Gamma_z/\chi$ in the calculations below.

In the following, we discuss different cases that depend  whether spin flip processes are allowed or not. Here it's convenient to define the spin flip probability $P_f$ by $\gamma_r=P_f\Gamma$. Consider $N\Gamma_z\gg \gamma_z$, based on Eq.~(\ref{eq:cavitygain}) and Eq.~(\ref{eq:spgain}), the metrological gain $\xi^{-2}=1/[N(\Delta\phi)^2]$ can be calculated as follows,
\begin{equation}
    \xi^2\approx \frac{1+2N\Gamma_zt_0}{(N\chi t_0)^2}+\frac{8}{3}\gamma_rt_0.
    \label{eq:withflip}
\end{equation}
Furthermore if we assume $N\Gamma_zt_0>1$ to reach optimal metrological gain, we can minimize the last two terms in Eq.~(\ref{eq:withflip}), which leads to $\xi^2_{\mathrm{opt}}=\sqrt{64P_f/3NC'}$ at $(\chi t_0)_{\mathrm{opt}}=\sqrt{3C'd^2/4NP_f}$. This assumption is valid if $(4P_f/3C'N)^{1/4}\ll d \ll 1$, and the optimal metrological gain is independent of the choice of $d$ in this regime. This result agrees with the calculation in Ref.~\cite{Davis2016}. Therefore, if the spin flip processes are allowed, the optimal sensitivity of gravimetry would be $\Delta g/g\propto N^{-3/4}$.

In the case of $P_f=0$, the metrological gain $\xi^{-2}$ now becomes
\begin{equation}
    \xi^2\approx \frac{1+(2N\Gamma_z+3\gamma_z)t_0}{(N\chi t_0)^2}+\frac{1}{N}+\frac{1}{2}(\chi t_0)^2.
    \label{eq:withoutflip}
\end{equation}
Based on Eq.~(\ref{eq:cooperativity}), we can first minimize the sum $2N\Gamma_z+3\gamma_z$ by $(2N\Gamma_z+3\gamma_z)_{\mathrm{min}}=4\chi\sqrt{3N/C'}$. This minimum can be reached by choosing $d=\sqrt{3/NC'}$. Similarly we assume $2N\Gamma_zt_0+3\gamma_zt_0>1$ to reach optimal metrological gain, so the minimization of Eq.~(\ref{eq:withoutflip}) leads to $\xi^2_{\mathrm{opt}}=[1+3(6/C')^{1/3}]/N$, which can be achieved at $(\chi t_0)_{\mathrm{opt}}=(48/C')^{1/6}/\sqrt{N}$. This assumption is valid if $C'\ll 48$. For large $C'$, the effect of decoherence would be negligible, and one will get the optimal metrological gain for ideal implementation $\xi^2_{\mathrm{ideal}}=e/N$ \cite{Davis2016}, where $e$ is the base of natural logarithm. Therefore, if the spin flip processes are forbidden, the optimal sensitivity of gravimetry would be $\Delta g/g\propto 1/N$.

\subsection{Inhomogeneous atom-light couplings}
Now we discuss the inhomogeneous OAT interactions, which are  described by the following Hamiltonian,
\begin{equation}
    H=\sum_{nm}\chi_{nm}S^z_nS^z_m,
\end{equation}
where $\chi_{nm}=\chi_{mn}$. Obviously we find $(\Delta S^y)^2_{\phi\rightarrow 0}=N/4$, which is not affected by the inhomogeneities. So we only need to focus on the calculation of $(\partial_{\phi}\langle S^y\rangle)_{\phi\rightarrow 0}$. Notice that
\begin{equation}
    \begin{aligned}
    (\partial_{\phi}\langle S^y\rangle)_{\phi\rightarrow 0}&=-i\langle \hat{x}|[S^y,U^{\dag}S^yU]|\hat{x}\rangle\\
    &=\frac{i}{4}\sum_{jk}\bigg[\langle \hat{x}|[\sigma_j^{+},U^{\dag}\sigma_k^{+}U]|\hat{x}\rangle-\langle \hat{x}|[\sigma_j^{-},U^{\dag}\sigma_k^{+}U]|\hat{x}\rangle-\langle \hat{x}|[\sigma_j^{+},U^{\dag}\sigma_k^{-}U]|\hat{x}\rangle+\langle \hat{x}|[\sigma_j^{-},U^{\dag}\sigma_k^{-}U]|\hat{x}\rangle\bigg],\\
    \end{aligned}
\end{equation}
where $|\hat{x}\rangle$ is a shorthand notation for the state $|S^x=N/2\rangle$, and $U=\exp[-it\sum_{nm}\chi_{nm}S^z_nS^z_m]$ is the time evolution operator of inhomogeneous OAT interactions. Here we elaborate the calculation of the first term $\langle \hat{x}|[\sigma_j^{+},U^{\dag}\sigma_k^{+}U]|\hat{x}\rangle$. Using the fact that the $|\hat{x}\rangle$ state can be expressed in the following form,
\begin{equation}
    |\hat{x}\rangle=\frac{1}{\sqrt{2^N}}\sum_{\sigma_1,\cdots,\sigma_N=\pm 1}|\sigma_1,\cdots,\sigma_N\rangle,
\end{equation}
where $\sigma_j=1/-1$ means the $j$-th atom in $|\uparrow\rangle/|\downarrow\rangle$ state, we have
\begin{equation}
    \begin{gathered}
    \sigma_k^{+}U|\hat{x}\rangle=\frac{1}{\sqrt{2^N}}\sum_{\begin{subarray}{c}\sigma_1,\cdots,\sigma_{k-1},\\\sigma_{k+1},\cdots,\sigma_N=\pm 1\end{subarray}}\exp\bigg[-i\frac{t}{4}\sum_{mn}^{(k)}\chi_{mn}\sigma_m\sigma_n-i\frac{t}{4}\chi_{kk}+i\frac{t}{2}\sum^{(k)}_m\chi_{mk}\sigma_m\bigg]|\sigma_1,\cdots,\sigma_k=+1,\cdots,\sigma_N\rangle,\\
    U\sigma_j^{-}|\hat{x}\rangle=\frac{1}{\sqrt{2^N}}\sum_{\begin{subarray}{c}\sigma_1,\cdots,\sigma_{j-1},\\\sigma_{j+1},\cdots,\sigma_N=\pm 1\end{subarray}}\exp\bigg[-i\frac{t}{4}\sum_{mn}^{(j)}\chi_{mn}\sigma_m\sigma_n-i\frac{t}{4}\chi_{jj}+i\frac{t}{2}\sum^{(j)}_m\chi_{mj}\sigma_m\bigg]|\sigma_1,\cdots,\sigma_j=-1,\cdots,\sigma_N\rangle,\\
    U\sigma_j^{+}|\hat{x}\rangle=\frac{1}{\sqrt{2^N}}\sum_{\begin{subarray}{c}\sigma_1,\cdots,\sigma_{j-1},\\\sigma_{j+1},\cdots,\sigma_N=\pm 1\end{subarray}}\exp\bigg[-i\frac{t}{4}\sum_{mn}^{(j)}\chi_{mn}\sigma_m\sigma_n-i\frac{t}{4}\chi_{jj}-i\frac{t}{2}\sum^{(j)}_m\chi_{mj}\sigma_m\bigg]|\sigma_1,\cdots,\sigma_j=+1,\cdots,\sigma_N\rangle,\\
    \sigma_k^{-}U|\hat{x}\rangle=\frac{1}{\sqrt{2^N}}\sum_{\begin{subarray}{c}\sigma_1,\cdots,\sigma_{k-1},\\\sigma_{k+1},\cdots,\sigma_N=\pm 1\end{subarray}}\exp\bigg[-i\frac{t}{4}\sum_{mn}^{(k)}\chi_{mn}\sigma_m\sigma_n-i\frac{t}{4}\chi_{kk}-i\frac{t}{2}\sum^{(k)}_m\chi_{mk}\sigma_m\bigg]|\sigma_1,\cdots,\sigma_k=-1,\cdots,\sigma_N\rangle,\\
    \end{gathered}
    \label{eq:resi}
\end{equation}
where $\sum^{(k)}_m$ means summation without the term $m=k$. Based on Eq.~(\ref{eq:resi}), we get (assuming $j\neq k$)
\begin{equation}
    \begin{aligned}
    \langle \hat{x}|[\sigma_j^{+},U^{\dag}\sigma_k^{+}U]|\hat{x}\rangle&=\langle \hat{x}|\sigma_j^{+}U^{\dag}\sigma_k^{+}U|\hat{x}\rangle-\langle \hat{x}|U^{\dag}\sigma_k^{+}U\sigma_j^{+}|\hat{x}\rangle\\
    &=\frac{1}{4}\mathrm{e}^{-it\chi_{jk}}\prod_{l}^{(j,k)}\cos(\chi_{kl}t)-\frac{1}{4}\mathrm{e}^{it\chi_{jk}}\prod_{l}^{(j,k)}\cos(\chi_{kl}t)\\
    &=-\frac{i}{2}\sin(\chi_{jk}t)\prod_{l}^{(j,k)}\cos(\chi_{kl}t).\\
    \end{aligned}
\end{equation}
It is easy to check that $\langle \hat{x}|[\sigma_j^{+},U^{\dag}\sigma_k^{+}U]|\hat{x}\rangle=0$ for $j=k$. Similar analysis can also apply to the other three terms based on Eq.~(\ref{eq:resi}) and  finally arrive to the following expression:
\begin{equation}
    (\partial_{\phi}\langle S^y\rangle)_{\phi\rightarrow 0}=\frac{1}{2}\sum_{j<k}\sin(\chi_{jk}t)\bigg[\prod_{l}^{(j,k)}\cos(\chi_{jl}t)+\prod_{l}^{(j,k)}\cos(\chi_{kl}t)\bigg].
\end{equation} 
For the times at which squeezing takes place, to an excellent approximation $\chi_{jk}t_0\ll1$ and therefore $(\partial_{\phi}\langle S^y\rangle)_{\phi\rightarrow 0}\sim  \sum_{j<k}\chi_{jk}t_0$. If the average is carried out over the full lattice array, we have $(\partial_{\phi}\langle S^y\rangle)_{\phi\rightarrow 0} \sim \frac{1}{2} N(N-1)\chi t_0$, and therefore inhomogeneities only give small higher order corrections. However, for applications such as measurements of short-range forces, that require the atoms to be placed  at local region of the lattice, then in this situation inhomogeneities does not average out in a single realization and can give rise to important systematic errors that will need to be accounted for. In this situation, operating with homogeneous couplings by satisfying the magic lattice condition (\ref{magic}) can provide an important measurement advantage. 

\subsection{Single-particle dephasing during interrogation time}
As pointed out by Ref.~\cite{Huelga1997}, single-particle dephasing imposes severe restrictions when operating with entangled states given their fragility to it. Here we estimate the longest interrogation time at which our protocol can still give a quantum advantage over uncorrelated states when the system is subjected to single-particle dephasing at a rate $\tilde{\gamma}_z$.
To simplify the analysis, instead of direct calculation of the twisting echo, we consider a spin squeezed state generated by OAT Hamiltonian. The master equation for interrogation takes the following form,
\begin{equation}
    \frac{\mathrm{d}}{\mathrm{d}t}\rho=i\bigg[\frac{\tilde{\delta}}{2}\sum_j\sigma^z_j,\rho\bigg]+\frac{\tilde{\gamma}_z}{4}\sum_j\bigg[\sigma^z_j\rho\sigma^z_j-\rho\bigg]
    \label{eq:dep}
\end{equation}
where $\tilde{\delta}=(\omega_{\mathrm{R}}-\omega_{\mathrm{MW}}-Mga_lr/\hbar)\times 2m_R$ in our case, and the system evolves under the Eq.~(\ref{eq:dep}) to accumulate a phase $\phi=\tilde\delta\tau$, where $\tau$ is the interrogation time. The measurement uncertainty on  the accumulated phase $\phi$ is given by
\begin{equation}
    \Delta\phi=\frac{\Delta S^y}{|\partial_{\phi}\langle S^y\rangle|}\bigg|_{\phi\rightarrow 0}.
\end{equation}
Therefore, the corresponding sensitivity for gravity measurements,  $\Delta g/g$, is given by
\begin{equation}
    \frac{\Delta g}{g}=\frac{\Delta\phi}{g|\partial_{g}\phi|}\sqrt{\frac{\tau}{T}}=\frac{\Delta\phi}{\omega_g\sqrt{\tau T}},
    \label{eq:sen}
\end{equation}
where $\omega_g=2Mga_lrm_R/\hbar$, and $T$ is the total averaging time.

The Heisenberg equation of motion that follow from Eq.~(\ref{eq:dep}) are the following  
\begin{equation}
    \begin{gathered}
        \frac{\mathrm{d}}{\mathrm{d}t}\langle\sigma_1^{+}\rangle=(-i\tilde\delta-\tilde{\gamma}_z/2)\langle\sigma_1^{+}\rangle,\\
        \frac{\mathrm{d}}{\mathrm{d}t}\langle\sigma_1^{+}\sigma_2^{-}\rangle=-\tilde{\gamma}_z\langle\sigma_1^{+}\sigma_2^{-}\rangle,\\
        \frac{\mathrm{d}}{\mathrm{d}t}\langle\sigma_1^{+}\sigma_2^{+}\rangle=(-2i\tilde\delta-\tilde{\gamma}_z)\langle\sigma_1^{+}\sigma_2^{+}\rangle.\\
    \end{gathered}
\end{equation}
For the spin coherent state along $\hat{x}$ direction, the initial condition for interrogation would be $\langle\sigma_1^{+}\rangle_{t=0}=1/2$, $\langle\sigma_1^{+}\sigma_2^{-}\rangle_{t=0}=1/4$, and $\langle\sigma_1^{+}\sigma_2^{+}\rangle_{t=0}=1/4$, which gives
\begin{equation}
    \langle S^y\rangle=N\mathrm{Im}\langle\sigma_1^{+}\rangle=-\frac{N}{2}\mathrm{e}^{-\tilde{\gamma}_z\tau/2}\sin\phi,
\end{equation}
\begin{equation}
    \begin{aligned}
    \langle S^yS^y\rangle&=\frac{N}{4}+\frac{N(N-1)}{2}\mathrm{Re}\bigg[\langle\sigma_1^{+}\sigma_2^{-}\rangle-\langle\sigma_1^{+}\sigma_2^{+}\rangle\bigg]\\
    &=\frac{N}{4}+\frac{N(N-1)}{8}\mathrm{e}^{-\tilde{\gamma}_z\tau}[1-\cos(2\phi)].\\
    \end{aligned}
\end{equation}
These results lead to  the following  sensitivity for gravity measurements 
\begin{equation}
    \frac{\Delta g}{g}=\frac{1}{\omega_g\sqrt{NT}}\sqrt{\frac{\mathrm{e}^{\tilde{\gamma}_z\tau}}{\tau}}.
    \label{eq:un}
\end{equation}
If instead a spin squeezed state generated by one-axis twisting Hamiltonian $H_{\mathrm{OAT}}=\chi S^zS^z$, followed by an additional rotation about $\hat{x}$ direction by an angle $\varphi$, is injected, the initial conditions for phase interrogation are
\begin{equation}
    \begin{gathered}
    \langle\sigma_1^{+}\rangle_{t=0}=\frac{1}{2}\cos^{N-1}(\chi t_0),\\
    \langle\sigma_1^{+}\sigma_2^{-}\rangle_{t=0}=\frac{1}{8}\bigg(1+\cos^{N-2}(2\chi t_0)\bigg)+\frac{1}{8}\cos^2\varphi\bigg(1-\cos^{N-2}(2\chi t_0)\bigg)-\frac{1}{2}\cos\varphi\sin\varphi\cos^{N-2}(\chi t_0)\sin(\chi t_0),\\
    \langle\sigma_1^{+}\sigma_2^{+}\rangle_{t=0}=\frac{1}{8}\bigg(1+\cos^{N-2}(2\chi t_0)\bigg)-\frac{1}{8}\cos^2\varphi\bigg(1-\cos^{N-2}(2\chi t_0)\bigg)+\frac{1}{2}\cos\varphi\sin\varphi\cos^{N-2}(\chi t_0)\sin(\chi t_0).\\
    \end{gathered}
\end{equation}
So we get
\begin{equation}
    \partial_{\phi}\langle S^y\rangle_{\phi\rightarrow 0}=-\frac{N}{2}\mathrm{e}^{-\tilde{\gamma}_z\tau/2}\cos^{N-1}(\chi t_0),
\end{equation}
\begin{equation}
    \langle S^yS^y\rangle_{\phi\rightarrow 0}=\frac{N}{4}+N(N-1)\mathrm{e}^{-\tilde{\gamma}_z\tau}\bigg[\frac{1}{8}\cos^2\varphi\bigg(1-\cos^{N-2}(2\chi t_0)\bigg)-\frac{1}{2}\cos\varphi\sin\varphi\cos^{N-2}(\chi t_0)\sin(\chi t_0)\bigg].
\end{equation}
Using a Ramsey spin squeezing parameter defined as  $\xi^2=\min_{\varphi}N(\Delta S_{\varphi}^{\perp})^2/|\langle\mathbf{S}\rangle|^2$, after  minimizing  with respect to rotation angle $\varphi$  one can calculate the following  sensitivity of this state for gravity measurements,
\begin{equation}
    \frac{\Delta g}{g}\approx\frac{1}{\omega_g\sqrt{NT}}\sqrt{\frac{\mathrm{e}^{\tilde{\gamma}_z\tau}-1+\xi^2}{\tau}}.
    \label{eq:en}
\end{equation}

We will assume that $1/\tilde{\gamma}_z$ can be set to be $100$~s in state-of-the-art experiments. In this case for unentangled initial states [see Eq.~(\ref{eq:un})], an interrogation time $\tau$ of the order of $10$~s is possible, if other technical noises do not impose further constraints at this time scale. For the spin squeezed states [see Eq.~(\ref{eq:en})] with $20$~dB metrological gain discussed in the main text, one would need to reduce the interrogation time $\tau \sim 1$~s in order to retain the quantum advantage of the initial state. In this case, our protocol not only reduces the required averaging time by a factor of $10$, but also increases the measurement bandwidth of time-varying signals by a factor of $10$, compared to unentangled lattice-based interferometers.